\documentclass{article} % For LaTeX2e
\usepackage{colm2024_conference}

\usepackage{microtype}
\usepackage{hyperref}
\usepackage{url}
\usepackage{booktabs}
\usepackage{amsmath}
\usepackage{algorithm}
\usepackage{algpseudocode}
\usepackage{multirow}
\usepackage{multicol}
\usepackage{hyperref}

\usepackage{arydshln}
\usepackage{booktabs}
\usepackage{stfloats}
\usepackage{graphicx}

\usepackage{titlesec}
\usepackage{booktabs}

\usepackage{adjustbox}
\usepackage{changepage} % 调整页面宽度
\usepackage{lipsum} % 示例文本

\usepackage{wrapfig}
\usepackage{lipsum}

\title{FuncEvalGMN: Evaluating Functional Correctness of SQL via Graph Matching Network}

\author{
Yi Zhan,  Dongchi Huang, Han Weng, Guifeng Wang, Jiajun Xie, Yu Tian, Boyi Liu, Yang Sun \\
ByteDance Inc.\\
}

\begin{document}

\maketitle

\begin{abstract}

In this paper, we propose a novel graph-based methodology to evaluate the functional correctness of SQL generation. Conventional metrics for assessing SQL code generation, such as matching-based and execution-based methods (e.g., exact set match and execution accuracy), are subject to two primary limitations.  The matching-based method does not effectively assess functional correctness, as different SQL queries may possess identical functionalities. The execution-based method is susceptible to producing false positive samples in evaluations. Our proposed evaluation method, \texttt{FuncEvalGMN}, does not depend on the sufficient preparation of the test data, and it enables  precise testing of the functional correctness of the code. Firstly, we parse SQL using a relational operator tree (ROT) called \textit{RelNode}, which contains rich semantic information from the perspective of logical execution. Then, we introduce a GNN-based approach for predicting the functional correctness of generated SQL. This approach incorporates global positional embeddings to address the limitations with the loss of topological information in conventional graph matching frameworks. As an auxiliary contribution, we propose a rule-based matching algorithm, RelNode Partial Matching (\texttt{RelPM}) as a baseline. Finally, we contribute a dataset, \texttt{Spider-Pair} with a training set and two testing sets, each comprising pairs of SQL codes to simulate various SQL code evaluation scenarios. \footnote{Our data and code are released at \url{https://github.com/hggforget/NL2SQL_partial_matching}}.
\end{abstract}

\section{Introduction}
Automatic code generation evaluation holds significant importance and broad prospects in the fields of natural language processing 
and software engineering.
%
% With the rapid development of Large Language Models (LLMs), code generation technology has gained widespread attention.
Owing to the powerful capabilities of Large Language Models (LLMs), there has been a marked surge in attention towards code generation directly from natural language prompts.
% %
% It has reduced the cost of program development and driven incredible progress in the field of software engineering.
% %
However, evaluating the performance of different code generation models remains challenging \citep{xu2022systematic}, which hinders the development of advanced code generation technologies. 
Thus, it is imperative to develop more precise and reliable evaluation methods.

In code generation tasks, text-to-SQL is a specific category that aims to interpret text semantics to predict accurate SQL queries \citep{yu2018spider}. In essence, there are three core types of metrics utilized for the evaluation of SQL. The most commonly used one is Execution Accuracy, which entails the comparison of the denotations of the predicted with the gold SQL based on executions. 
% YS: Not useful.
% SQL is a declarative language, that focuses on the desired data operation outcomes rather than the program's behaviors and state changes. 
Unlike PassRatio and Pass@k used in imperative languages, which are continuous by assessing code's pass rate on test cases, Execution Accuracy can only get a binary score (correct or incorrect), failing to reflect partial correctness, and is prone to high false positives when the test data is not comprehensive \cite{zhong2020semantic}. Furthermore, it incurs a significant setup cost for the test environment and computational load.
% metric 2
Another prevalent evaluation metric is matching-based, such as BLEU \citep{papineni2002bleu}, which relies on counting overlapping n-grams between the generated code and reference code \citep{zhou2023codebertscore}. 
However, they focus on basic and lexical-level features, failing to fully capture the wide range of functionally equivalent program variations \citep{dong2023codescore}. 
% metric 3
Conversely, an evaluation approach based on pre-trained models \citep{dong2023codescore, zhou2023codebertscore} utilizes contextual embeddings to assess similarity scores for candidate codes, transcending mere syntactic and textual attributes. Since the correctness of the code's functionality is inextricably linked to the code's syntactic structure and execution logic, constructing the code into a program graph can capture more semantic information than processing the code into a sequence of tokens \citep{mi2023graph}. Furthermore, within the realm of SQL, researchers transform the query equivalence problem into a constraint satisfaction problem and utilize a generic verifier to determine query equivalence \citep{zhou2022spes, 10.1145/3514221.3526125, chu2018axiomatic}. The technique consists of deriving symbolic representations of queries and proving their equivalence by determining query containment relationships between symbolic representations. However, there are some SQL keywords that cannot be converted into equivalent symbolic representations, which prevents this approach from being applied to all query.
% However, these models transform code into a sequence of tokens, impeding the ability to accurately capture the code's internal structure and hierarchy. Thus, its ability to evaluate codes is underdeveloped.
% \wh{As a result, the existing metrics for evaluating SQL code are notably underdeveloped.}

For the above reasons, we developed a novel metric, \texttt{FuncEvalGMN}, based on graph similarity.
% %%
% \wh{the relation between graph matching and SQL evaluation}
% To address these challenges, we introduce a novel graph-based evaluation metric, \texttt{FuncEvalGMN}, that excels at capturing the complex syntactic structure and semantic information of the code, significantly improving the evaluation of functional correctness. 
% \sy{To address the above challenges, we developed a novel metric based on graph similarity. The latter part has significant problems.}
Firstly, we convert SQL into a Relational Operator Tree (ROT), called \textit{RelNode}, that abstracts SQL's specific syntax and represents its logic execution plan. Then a comprehensive program graph is constructed by incorporating logical and data flows, which monitor the execution states and data usage in SQL. We utilize the graph similarity as the metric to evaluate the functional correctness of the generated SQLs by Graph Matching Network (GMN) \citep{li2019graph}. Innovatively, we introduce a global positional embedding in GMN's cross-attention computations, enhancing the model's ability to capture cross-graph structural details. In addition, we have constructed a matching-based approach, RelNode Partial Matching (\texttt{RelPM}) as one of our baselines. Furthermore, we created a dataset, \texttt{Spider-Pair}, based on Spider \citep{yu2018spider}, where each data entry includes a prompt constructed from table schemas and questions, a SQL pair (reference and generated SQL), and the functional correctness of generated SQLs. Specifically, it has one training and two testing sets. On our dataset, \texttt{RelPM} achieves AUCs of 78.17\% and 62.37\% on two test sets, significantly surpassing other matching-based metrics, while \texttt{FunEvalGMN} with 94.32\% and 84.8\% AUCs, achieves state-of-the-art performance among all methods. In addition, the introduction of PE improves up to 4\% compared to the original GMN, demonstrating its effectiveness in strengthening the performance of matching algorithms.

The main contributions of this paper are as follows:
(a) We propose a generic distance metric based on embedding the program graph for evaluating the functional correctness of SQL generation. (b) We introduce a global positional encoding mechanism in the cross-attention computations of GMN, enhancing the representation capability with cross-graph structural information. (c) We develop a custom matching algorithm, \texttt{RelPM}, that can provide a comprehensive score by adjusting the weights of parent and child nodes. (d) We build a dataset, \texttt{Spider-Pair}, filling gap in evaluating SQL generation.

% (a) We propose a novel graph-based approach to evaluate the functional correctness of SQL generation. A generic distance metric for evaluation of the semantic similarity based on embedding the program graph. (b) We introduce an innovative positional encoding mechanism in the cross-attention computations of GMN, enhancing the representation capability with cross-graph structural information. (c) We develop a custom matching algorithm, \texttt{RelPM}, that can provide a comprehensive score by adjusting the weights of parent and child nodes and paying more attention to the important clauses and key nodes. (d) We build a dataset, \texttt{Spider-Pair}, filling gap in evaluating SQL generation. This dataset not only includes generation by LLMs but also incorporates data augmentation through SQL equivalence rewriting. 

\section{Related Work}

% This section is comprised of three parts. In the first part, we introduce the existing code evaluation metrics research in Section \ref{cem}. Then, we present the Graph-based code representation in Section \ref{graph code representation}. Finally, we focus on the latest Graph Matching Neural Networks in Section \ref{gmnn}.

\textbf{Code Evaluation Metrics:} \label{cem} The execution-based method becomes prevalent code evaluation harness \citet{chen2021evaluating, xu2022systematic, kulal2019spoc}. Execution Accuracy compares the denotations of the ground truth and predicted SQL via execution on a database \citep{yu2018spider}. This approach requires databases that cover data distributions to prevent false positives, whereby functionally different SQL happens to be incorrectly identified as correct. Test suite accuracy (\citet{zhong2020semantic}) involves distilling datasets to create compact, high-coverage database test suites for effective SQL semantic accuracy testing.

On the other hand, Matching-based evaluation metrics assess surface-form differences in code by considering the code’s syntactic rules and structured nature, CrystalBLEU \citep{eghbali2022crystalbleu} evaluates predicted codes quality by empirically eliminating the most common n-grams lacking semantic significance in code, while CodeBLEU \citep{ren2020codebleu} incorporates syntactic structure and semantic information of code through Abstract Syntax Trees (AST) and data flow analysis. However, these methods fail to correctly evaluate the code snippets that are semantically equivalent but lexically diverse, causing serious {\it false negatives}.

% In SQL programming, evaluation of the functional correctness of the predicted SQL poses a challenging problem, Execution Accuracy compares the denotations of the ground truth and predicted SQL via execution on a database \citep{yu2018spider}. This approach requires databases that cover data distributions to prevent false positives, whereby functionally different SQL happens to be incorrectly identified as correct. \citet{zhong2020semantic} involves distilling datasets to create compact, high-coverage database test suites for effective SQL semantic accuracy testing.

More recently, code representation with {\it pretrained models}, exhibits excellent code comprehension and significantly aid in code evaluation. CodeBERTScore \citep{zhou2023codebertscore} encodes codes into contextual embeddings using CodeBERT \citep{feng2020codebert} and calculates their vector similarity. CodeScore \citep{dong2023codescore} proposes a unified code generation learning framework for pre-trained models to learn code execution.

\textbf{Graph-based Code Representation:} \label{graph code representation} Representing source code without loss of important information is an active area of research for code analysis in various software engineering tasks such as code completion \citep{wang2021code, liu2022unified}, code clone detection \citep{fang2020functional, yu2023graph} and code summarization \citep{tang2022ast, tang2021ast}. AST is a tree-shaped structure representing the abstract syntactic structure of code. It abstracts away specific details from the real syntax, focusing only on the structural and content-related aspects. To augment original ASTs, \citet{mi2023graph, allamanis2018learning, wang2020detecting} integrate explicit semantic edges, such as data flow and control dependencies, constructing a comprehensive program graph containing both syntactic and semantic information. Additionally, three primary graphical representations in program analysis are Control Flow Graphs (CFGs) \citep{cota1994control}, Data Flow Graphs (DFGs) \citep{orailoglu1986dataflowgraphrepresentation}, and Program Dependence Graphs (PDGs) \citep{ottenstein1984programdependencegraph}. 
% CFGs map out execution paths, highlighting the control flow through program statements, essential for understanding conditional and looping structures, while DFGs trace how data moves and changes, using nodes for variables or operations and edges for data flow directions. PDGs combine CFGs and DFGs, making them valuable for detailed program analysis by providing a comprehensive view of a program's structure and behavior. PS-by-Yang: 
\citet{fang2020functional, shi2023coss} use CFGs to represent codes, as they address the limitations of ASTs in capturing semantic features and the coarse granularity of PDGs in detailed code analysis. However, SQL lacks structures such as CFGs, DFGs, and PDGs. Contemporary approaches primarily focus on extracting SQL's syntactic and structural information from ASTs \citep{cao2023heterogeneous, zhuo2021long, cao2023astormer}. To understand SQL queries' data flow and logic dependencies, the Relational Operator Tree (ROT) represents SQL as a sequence and hierarchy of relational operators \citep{cyganiak2005relational}, which depicts the logical execution plan of the queries. Our research is pioneering in combining ROT with logic and data flow to extract SQL syntax and semantics for code analysis.

\textbf{Graph Matching Neural Network:} \label{gmnn} The graph matching problem plays a pivotal role in various real-world applications, including source/binary code analysis \citep{yu2020codecmr, xu2017neural} and computer vision \citep{chen2021learning, liu2020graph}. 
One of the early approaches, SMPNN \citep{riba2018learning}, approached the problem by modeling similarity as a summation of node-node similarity scores. 
Following this, SIMGNN \citep{bai2019simgnn} proposed utilizing node-node similarity scores by extracting histogram features of nodes. 
In a significant development, GMN \citep{li2019graph} innovatively incorporated node-node similarity information into graph-level embedding via a cross-graph attention mechanism, enhancing graph characterization by considering node differences across two graphs. 
GraphSim \citep{bai2020learning} diverged from the graph-level representation approach, arguing that a single fixed-dimensional embedding per graph might not adequately represent graphs of varying sizes and link structures. 
Instead, it conducts operations directly on sets of node embeddings. 
Advancing in complexity and depth of graphs, MGNN \citep{ling2021multilevel} introduced a multi-view matching function, comparing contextual node embeddings of one graph with the graph-level embedding of another, effectively capturing intricate cross-level interaction features. 
Focusing on model efficiency, EGSC \citep{qin2021slow} developed a collaborative attention-based feature fusion network, employing knowledge distillation in the fusion model to expedite the inference process. Concurrently, ERIC \citep{zhuo2022efficient} introduced a powerful alignment regularization technique, applying node-graph correspondence constraints on the GNN encoder to reduce computational cost.

% \textnormal{The tree on the left is an Abstract Syntax Tree (AST), which abstracts SQL purely from a syntactic perspective, making it difficult to reflect the execution logic and data flow process of SQL. In contrast, \textit{RelNode} analyzes SQL from the standpoint of a logical execution plan. Each of its subtrees signifies an execution step and meticulously outlines parameter utilization, providing richer semantic information.}

\begin{figure}
    \centering
    \includegraphics[width=1\linewidth, scale=1.00]{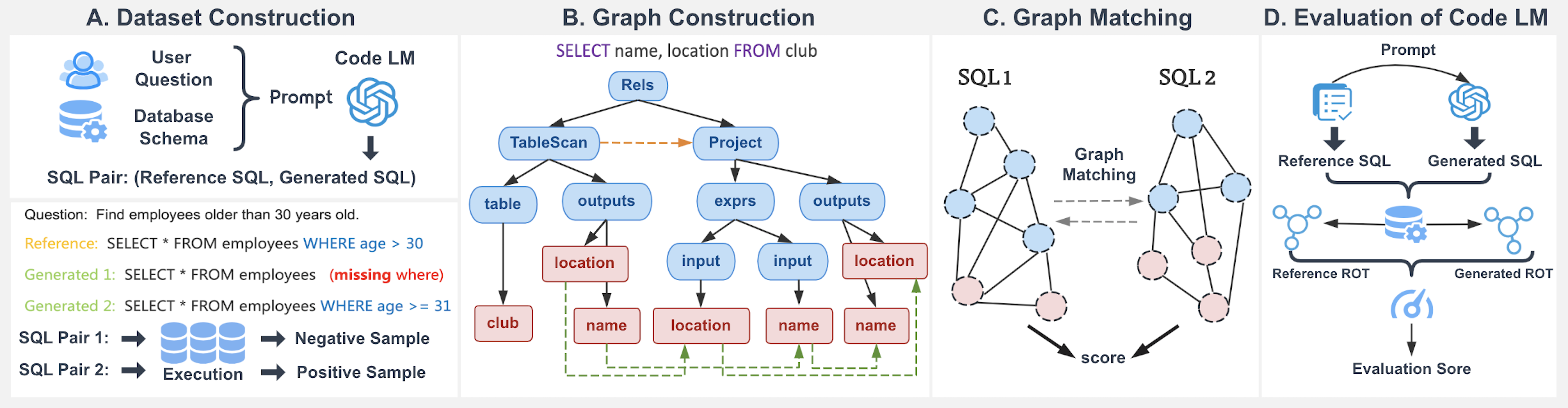}
    \caption{Overview of FuncGMN approach}
    \label{relnode}
\end{figure}

\section{RelNode Partial Matching}

\label{sec: relnode}
% \wh{why relnode? add reason first}
% Analyzing code as natural language overlooks its rich syntactic structure \cite{ren2020codebleu, eghbali2022crystalbleu}, whereas graph-based methods can offer an effective way to abstract the syntactic structure of code without losing important information \citep{mi2023graph, allamanis2018learning, wang2020detecting}.
\citet{ren2020codebleu, eghbali2022crystalbleu} suggest that merely interpreting code as natural language for evaluation of model generation may neglect its intricate syntactic architecture, whereas graph-based methods can be effective to abstract the syntactic structure of code without losing important information \citep{mi2023graph, allamanis2018learning, wang2020detecting}. To this end, we convert the SQL query to a Relational Operator Tree (ROT), which represents SQL's logic execution plan using relational algebra. It effectively illustrates the structural order and hierarchical organization of relational operators such as join, selection, and projection \citep{cyganiak2005relational}. By leveraging Apache Calcite \citep{begoli2018apache}, an optimized ROT is obtained, called \textit{RelNode}\footnote{\url{https://github.com/apache/calcite/blob/calcite-1.36.0/core/src/main/java/org/apache/calcite/rel/RelNode.java}}. It refines the plan through operation reordering and redundant clause elimination. After optimization, the \textit{RelNodes} can uncover similar logical execution patterns beneath varied syntactic forms, enabling precise semantic equivalence comparisons between SQL statements (an example is shown in Appendix \ref{a:relnode}). We categorize the nodes of \textit{RelNode} into \textit{Computing Nodes} and \textit{Content Nodes} for convenience of representation. \textit{Computing Nodes} act as operators, forming the syntactic structure of SQL queries, while \textit{Content Nodes}, representing the query's parameter variables, serve as operands and are the leaf nodes of the tree.

\begin{minipage}[t]{.48\textwidth} % Left column
    % Algorithm 1
    \begin{algorithm}[H]
    \caption{ROT Node Comparison}
    \label{al1}
    \begin{algorithmic}[1]
    \Function{Calc}{$node$, $Node$}
    \If{$node.val == Node.val$}
    \State \Return 1
    \EndIf
    \State \Return 0
    \EndFunction
    \end{algorithmic}
    \end{algorithm}

    % Space between algorithms (optional, adjust as needed)
    % Algorithm 2
    \vspace{-0.2cm}

    \begin{algorithm}[H]
    \caption{RelPM Score}
    \label{al2}
    \begin{algorithmic}[1]
    \State Let $S$ be the source tree.
    \State Let $T$ be the target tree.
    \Procedure{RelPM}{S, T}
    \State $recall \gets \Call{NodeMatch}{S, T}$
    \State $precision \gets \Call{NodeMatch}{T, S}$
    \State $F_{\beta} = \frac{(1 + \beta^2) \times precision \times recall}{\beta^2 \times precision + recall}$
    \State \Return $F_{\beta}$
    \EndProcedure
    \end{algorithmic}
    \end{algorithm}
    
\end{minipage}%
\hfill
\begin{minipage}[t]{.5\textwidth} % Right column
    % Algorithm 3
    \begin{algorithm}[H]
    \caption{Node Match}
    \label{al3}
    \begin{algorithmic}[1]
    % \State Let $S$ be the source tree.
    % \State Let $T$ be the target tree.
    % \State Let $s$ be a node in the source tree.
    % \State Let $t$ be a node in the target tree.
    
    \Function{NodeMatch}{$S$, $T$}
    \State $m_{root} \gets \Call{Calc}{S, T}$
    \If{$S$ or $T$ is a leaf node}
    \State \Return $m_{root}$
    \Else
    \State $scores \gets \text{an empty list}$
    \For{each child $s$ in $S.children$}
    \State $m \gets 0$
    \For{each child $t$ in $T.children$}
        \State $m \gets \max(m, ($
        \State $\quad\quad(\Call{NodeMatch}{s, t}))$
    \EndFor
    \State $scores$.add($m$)
    \EndFor
     \State $m_{children} = \frac{\text{Sum}(scores)}{\text{Length}(scores)}$
    \State \Return $m_{root} * \alpha + (1-\alpha) * m_{children}$
    \EndIf
    \EndFunction
    \end{algorithmic}
    \end{algorithm}
\end{minipage}

% After abstracting the SQL query to the \textit{RelNode} format, we apply a simple partial matching method to serve as the benchmark for the matching-based evaluation.
% In the short, \texttt{\texttt{RelPM}} is a general matching algorithm calculates the matching nodes and their scores for each \textit{RelNode}. 
% The matching algorithm can be applied to any tree structure. 
% If the structure being matched is an AST, we call it an ASTPM. 
% It is suggested in \cite{ren2020codebleu}, that using a matching-based approach on AST between reference code and generated code leads to a better correlation of the metric with the code quality. Here, we present our rule-based matching method for the evaluation of generated SQL quality.

Upon converting the SQL query to the \textit{RelNode}, we establish a rule-based partial matching algorithm (\texttt{RelPM}) to serve as the benchmark for the matching-based evaluation. In short, \texttt{RelPM} is a general matching algorithm applied to the tree structure, calculating the score for every matching node. % 详细描述
% The partial matching process is divided into three parts: Node Matching, RelNode Scoring, and Similarity Evaluation. 
% For brevity, Node Matching and RelNode Scoring are combined. Algorithm \ref{rpm} displays the pseudocode of \texttt{RelPM}. 
% In the Node Matching Algorithm \ref{rpm}, each node's score is a weighted sum of its own and its children's scores, based on characteristic comparisons with a matched node, and the maximum scores of its potential matched children in the other tree, recursively calculated to the leaf nodes.
In the Node Matching Algorithm \ref{al3}, the score of each node is calculated as a weighted sum of the node's own score and the scores of its children, with weights assigned according to comparisons of their characteristics and Algorithm \ref{al1} is used to compare the characteristics of two nodes. The parameter $\alpha$ is a globally adjustable free parameter, subject to fine-tuning to ensure a balanced allocation of matching weights between the root node and its child nodes. Note that Algorithm \ref{al3} is asymmetric with respect to $S$ and $T$. In Algorithm \ref{al2}, we use an adjustable parameter \( \beta \) to calculate the weighted geometric mean to control the focus on the semantics completion of the source tree.

% Finally, using the reference tree as the source, the score is calculated as recall, and for the generated tree, the score is precision. 

\section{RelNode Graph Matching Network}
\label{sec: gmn}
% \wh{change subsubsection to paragraph,8 paragraph two much. 2 suggestions: 1, make the subsection a new method section since its our novel method indeed. previous break up 2, shorten this section. }
% As a method based on matching, \texttt{RelPM} cannot entirely avoid the effects associated with textual variations and differences in syntactic structure.
As a method based on matching, \texttt{RelPM} remains susceptible to recognize identical semantics of the SQL with syntactic structure change.
% Therefore, we propose a novel approach, \texttt{FuncEvalGMN}, based on the graph matching network, to further capture the functional correctness of different SQLs.
Therefore, we propose a novel approach, \texttt{FuncEvalGMN}, based on the graph matching network, to further capture the functional correctness of generated SQLs.

% In the \texttt{FuncEvalGMN}, we build a program graph based on \textit{RelNode}, which incorporates both data flows and logic flows. Subsequently, we embed the features and positions of the nodes.
% A graph-matching neural network then computes the embeddings of these graphs. 
% To evaluate the functional correctness of generated SQL, we employ a vector space similarity metric. Each of these steps will be detailed further.

\subsection{Program Graph Construction}

\begin{figure}[H]
    \centering
    \includegraphics[width=\linewidth, scale=1.00]{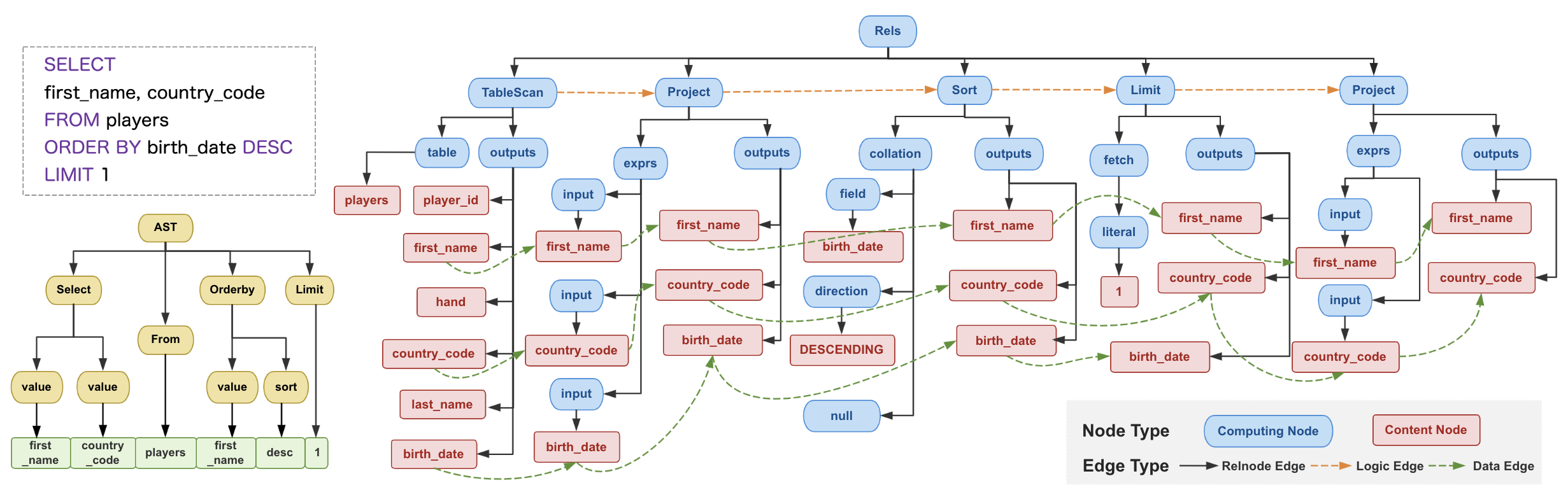}
    \caption{SQL Representations using \textit{RelNode} and AST. The AST purely abstracts SQL from a syntactic perspective, whereas \textit{RelNode} provides more semantic information from an execution standpoint. In \textit{RelNode}, orange edges represent logic flows, connected at the second layer of the \textit{RelNode}, indicating the logical sequence of clause execution. Green-colored data flows represents the pathways of data across various clauses, associating with nodes representing column names. }
    \label{relnode}
\end{figure}

Based on \textit{RelNode}, we integrate data and logic flows to thoroughly capture the syntactic and semantic information of SQL.  The logic and data flows enhance the ability of graphs to accurately analyze the complex interactions and dependencies within SQL subclauses. Finally, the SQL is represented as a program graph \( G = (V, E) \), where \( V \) are nodes from the \textit{RelNode}, \( E \) are edges including the edges of \textit{RelNode}, and data and control flow edges.

% For instance, examining the data flows of \textit{birth\_date}, we observe that it originates from the table and is then utilized as the key field in the \textit{Sort} operator. Following the execution of \textit{Limit}, \textit{birth\_date} is ultimately disregarded in the final \textit{Project} step. 

\subsection{Node Feature Embedding}

% As illustrated in Figure \ref{pipeline}(B), the heterogeneous program graph we constructed comprises two types of nodes: \textit{Computing Nodes} and \textit{Content Nodes}. \textit{Computing Nodes} constitute the primary syntactic structure of the AST while \textit{Content Nodes} are the leaf nodes and reflect parameter variables in SQL. We refer these information as the attribute features of the nodes. Nodes with same neighborhood tend to have similar representations. However, their functions differ due to their distinct hierarchical or subtree positions. Therefore, we introduce positional encoding to enhance their nodes.

% Additionally, in GNNs, nodes with the same neighborhood often have similar representations. However, their functions may differ due to unique hierarchical positions or locations within subtrees. To address this, we introduce positional encoding to augment the distinction of these nodes. In database systems, various relational algebra expressions can equivalently translate SQL for execution.

% In program graph, nodes are classified into \textit{Computing Nodes} and \textit{Content Nodes}. Given that they contain distinct types of information, we employ different methods to encode their features into vectors.

% \subsubsection{Computing Nodes}

\textbf{Computing Nodes:} In SQL, many equivalent syntactic structures can perform the same functional operations. For instance, a combination of \textit{ORDER BY} and \textit{LIMIT} indicate an equivalent \textit{MAX} or \textit{MIN} operation.
% Therefore, our approach will capture the overall semantic information of \textit{Computing Nodes}, rather than emphasizing their individual features. % Assuming there are $k$ distinct types in \textit{Computing Nodes}, each of them is represented by a $k$-dimensional vector \( X \in \mathbb{R}^k \). Then we map these nodes to initial node embeddings via Multilayer Perceptrons (MLPs):
% \[
% h_i^{(0)}=\quad\text{MLP}_\text{computing}{ ( \mathbf{x}_i)},\quad\forall i\in V
% \]
Assuming there are $k$ distinct types in \textit{Computing Nodes}, we compute the embeddings of each $\mathbf{x}_i \in [0, k-1]$ as
\[
h_i^{(0)}= \text{Embedding}{ ( \mathbf{x}_i)},
\]  where the superscript $(0)$ represents the initial state, which will be updated as message propagates through the graph neural network.

% we have designed a node encoder based on ASCII encoding, whose core components are 1D convolutional layers and residual blocks. % Due to the unambiguous of programming languages, 
\textbf{Content Nodes:} The \textit{Content Nodes} represents the query's parameter variables, which is also the precise database schema elements. A common challenge faced in text-to-SQL is that SQL queries, despite being structurally sound, may fail to execute as intended due to the misuse of parameter variables, such as table names and column names. The main reason is that word embedding models such as \textit{Word2Vec} \cite{mikolov2013efficient} and \textit{FastText} \cite{bojanowski2017enriching} is adopted for representing the semantics, with their tendency to represent two semantically similar but distinct entities (i.e., column name `kid' and `child') with similar embeddings. 
%For example, mistakenly changing the actual column `kid' to a non-existent column `child' in SQL can result in execution failure. Although word embedding models such as \textit{Word2Vec} \cite{mikolov2013efficient} and \textit{FastText} \cite{bojanowski2017enriching} are widely used for representing the semantics, they are unsuitable in our tack because their tendency to represent two semantically similar but distinct entities (i.e., column `kid' and `child') with similar embeddings. 
To address this challenge, a string-aware embedding method is introduced to enhance our model's understanding of different entities. Specifically, we first characterize each \textit{Content Node} as a string \( S = (s_1, s_2, \ldots, s_n) \) of fixed length \( n \), with each element is encoded as an ASCII value ranging from 0 to 127, it could be transformed into an n-dimensional vector \( X \in \mathbb{R}^{1 \times n} \). Then, we apply one-hot encoding with 128 dimensional size to each dimension, expanding the vector to \( X \in \mathbb{R}^{n \times 128} \). To enhance the feature difference among various nodes, a ResNet model \citep{he2016deep} is further adopted. Due to page limitation, please refer to Appendix \ref{aresnet} for the specific architecture.

% SQL with the same syntax could function incorrectly with misused parameter variables, such as table name and column name. Thus, for \textit{Content Nodes}, we focus on its string-level features. For example, `Kid' and `Child' represent entirely different entities. Word embedding models such as \textit{Word2Vec} \cite{mikolov2013efficient} and \textit{FastText} \cite{bojanowski2017enriching} are unsuitable in our tack since they will encode them into similar representations. To overcome this, we first encode these nodes using ASCII. As Figure \ref{resnet} shows, for a \textit{Content Node} characterized by a string \( S = (s_1, s_2, \ldots, s_n) \) of length \( n \), we transform it into an n-dimensional vector \( X \in \mathbb{R}^n \), representing each character by its ASCII value. Given the ASCII range of 0 to 127, we apply one-hot encoding to each dimension, expanding \( X \) to \( X \in \mathbb{R}^{n \times 128} \). Then, we enhance the feature difference among various nodes by a ResNet model \citep{he2016deep}, and the architectural diagram of the model is shown in Appendix \ref{aresnet}.

\subsection{Positional Embedding}

Nodes with the same neighborhood often exhibit similar representations. However, their actual functions might vary due to unique hierarchical positions or locations within subtrees. To address this, we incorporate positional encoding, specifically using the Random Walk Positional Encoding (RWPE) \citep{dwivedi2021graph}. It's defined using a k-step random walk:
\[ p_{i}^{\text{rw}} = [ T_{ii}, T_{ii}^2, \ldots, T_{ii}^k ] \in \mathbb{R}^k, \quad k \in [1, K] \]
where $T$ is the state transition matrix of the random walk, and \( T^k \) represents the $k$-th power of $T$. $A$ and $D$ represent the adjacency matrix and degree matrix of the graph. This method utilizes a simple random walk matrix approach, focusing solely on the probability of node \( i \) returning to itself, denoted by \( T_{ii} \). This strategy offers a distinct node representation, based on the premise that each node possesses a unique $k$-hop topological neighborhood, particularly when \( k \) is sufficiently large.

\begin{figure}[h]
  \centering
  \begin{minipage}[b]{0.5\textwidth}
    \centering
    \includegraphics[width=\textwidth]{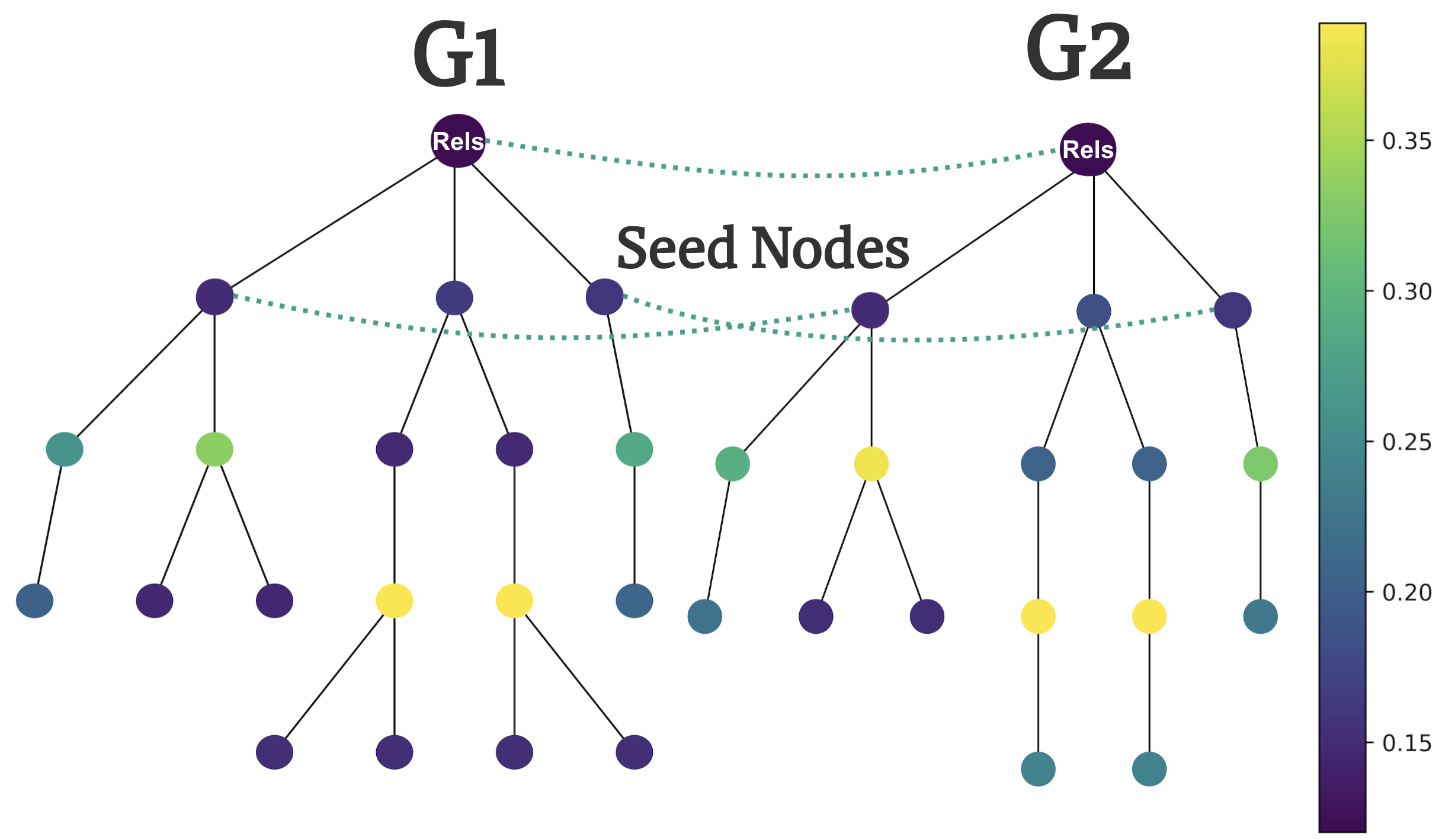} % 替换为第一张图片的文件名
    \caption{Positional Encoding.}
    \label{fig:image1}
  \end{minipage}
  \hfill
  \begin{minipage}[b]{0.45\textwidth}
    \centering
    \includegraphics[width=\textwidth]{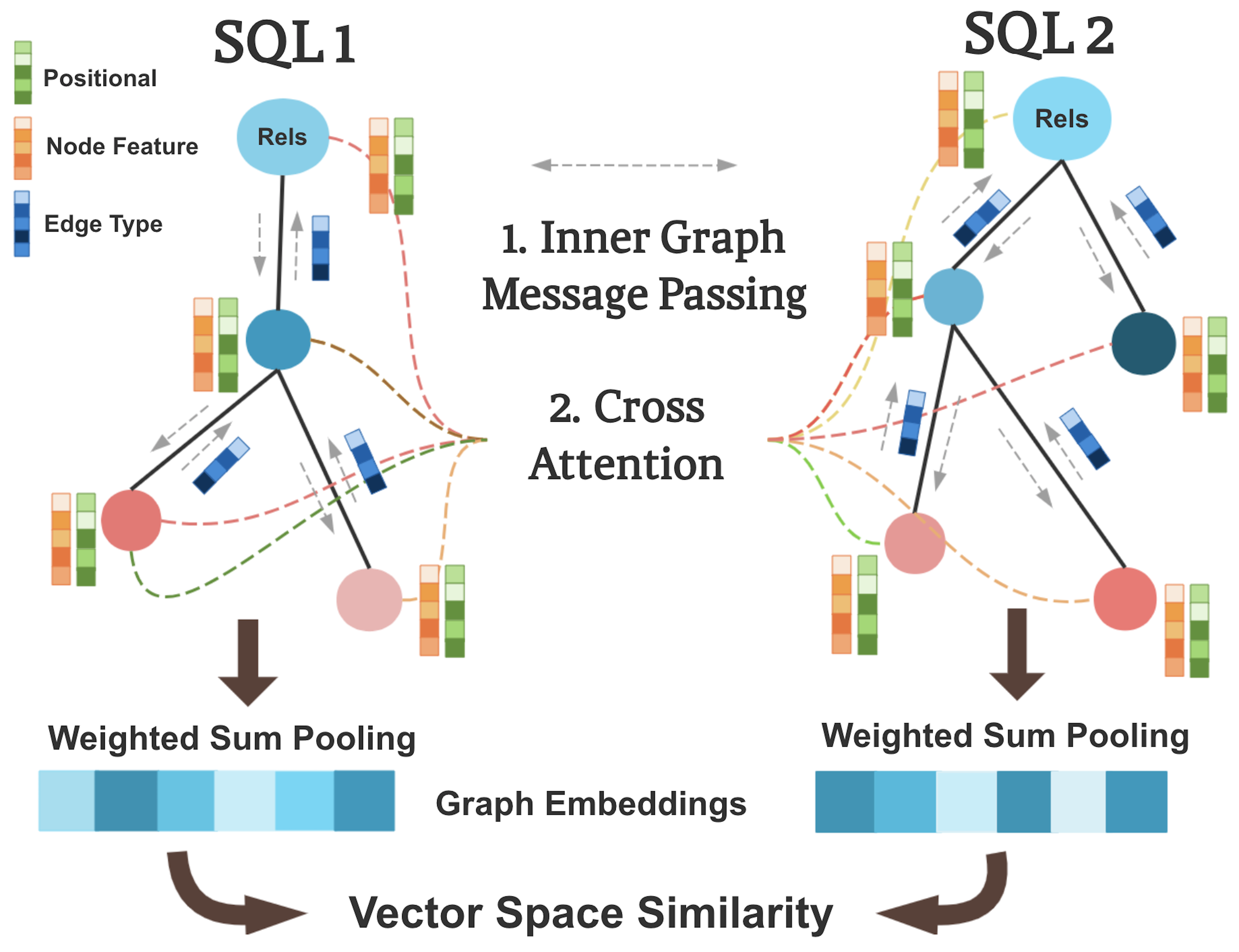} % 替换为第二张图片的文件名
    \caption{Graph Embedding.}
    \label{fig:image2}
  \end{minipage}
\end{figure}

% \begin{figure}
%     \centering
%     \includegraphics[width=0.5\linewidth,  scale=1.00]{pic/pe.png}
%     \caption{Positional Encoding in Merged Graph. \textnormal{We connect the root node with its isomorphic subtree to construct a merged graph and perform the random walk for 4 times. The node colors represent the probability of returning to themselves on average per walk, which highlights that nodes with similar substructures tend to have closer feature values.}}
%     \label{pe}
% \end{figure}

% \begin{figure}
%     \centering
%     \includegraphics[width=0.6\linewidth, scale=1]{pic/gmn_1.png}
%     \caption{Graph Embedding. \textnormal{After message passing, we use weighted sum pooling to get graphs' final representation, and finally calculate their vector space similarity.}}
%     \label{gmn}
% \end{figure}

 Two methods to compute node positional encodings for the graph-pair are proposed. The first method calculates the positional encodings of nodes within each graph separately, while the second method, as Figure \ref{fig:image1} shows, connects the two graphs using \textit{Rels} node and predefined \textit{seed} nodes. Then, we calculate the global positional encodings for nodes on the merged graph. Evidently, the latter provides a global positional information for nodes within and across graphs. \textit{Rels} node refers to the top nodes in each \textit{RelNodes}  and \textit{Seeds} nodes refer to nodes that match between two \textit{RelNodes} with their corresponding subtrees $S_1$ and $S_2$. Namely, $\texttt{RelPM}(S_1, S_2) = 1$. We restrict cross-graph edges to include root nodes of second-level operators representing subclauses, such as \textit{TableScan} and \textit{Sort}. In Figure \ref{fig:image1}, the node colors represent the average probability of returning to themselves after $K=4$ walks, which highlights that nodes with similar substructures tend to have closer feature values. Additionally, in merged graph, each node has more random walk paths to choose from, reducing the likelihood of returning to itself, so the node's hierarchical positions become more prominent. %We map these initial positional embeddings via Multilayer Perceptrons (MLPs):

\subsection{Graph Embedding}

Consider two graphs constructed from SQL, \(G_1 = (V, E)\) and \(G_2 = (V, E)\). 
% Each node \(v \in V\) having an associated characteristic embedding \(x_v\) and positional embedding \(p_v\), and each edge \((u, v) \in E\) is linked to a type vector \(e_{uv}\). 
We create a Graph Matching Network (GMN) based on \citet{li2019graph} to generate graph representations and assess their similarity. In this section, we concentrate on enhancing the GMN's message-passing mechanism, as shown in Figure \ref{fig:image2}. More details are deferred to Appendix \ref{argmn}.

\textbf{Inner-graph Message Passing}: In inner-graph message passing, for a node \(v\), its neighbor set is denoted as \(N(v)\). The message passed at step \(t\) is:
\[
m^{(t+1)}_v = \sum_{u \in N(v)} f_\mathrm{inner}(h^{(t)}_v, h^{(t)}_u, e_{uv}),
\] where \( m^{(t+1)}_v \) represents the message received by node \( v \) at setp \( t+1 \), \( h^{(t)}_v \) and \( h^{(t)}_u \) are the representation of node \( v \) and \( u \) in step \( t \). $e_{uv}$ is the representation of edge between $u$ and $v$. There are three types of edges, we compute the embeddings of each $\mathbf{e} \in [0, 2]$ as
\[
e_{uv}= \text{Embedding}{ ( \mathbf{e})},
\]

% Specifically, the message function \( f_\mathrm{inner} \) is an MLP in our model.

\textbf{Cross-graph Message Passing}: In cross-graph message passing, cross-attention is utilized to compute graph representation and hence capture the semantic similarity of two graphs jointly, rather than focusing on isolated features within a single graph. Specifically, for a node \(v\) in graph \(G_1\), we consider \(G_2(v)\) to represent the corresponding set of nodes in graph \(G_2\). Different from inner-graph message passing, the node representation $r_v$ here is a nonlinear combination of the node's feature embedding $x_v$ and the node's positional embedding $p_v$. Furthermore, the node's positional representation is updated in this process \citep{dwivedi2021graph}. % In step \(t\). 
On top of \citet{li2019graph}, which uses fully connected cross attention, our cross attention with merged-graph positional encoding gives more expressive node representation. For example,  different \textit{birth\_date} nodes, as shown in Figure \ref{relnode}, under different subtrees (under \textit{Sort} and \textit{Project}) can be effectively identified. %The $f_\mathrm{cross}$ function passes the difference between one node and the node of the other graph, which can be calculated by the following equations:
$\mu_v$ is cross-graph message between node $v$ to another graph, which is computed with cross attention as  

\[
\mu^{(t+1)}_v = \sum_{u \in G_2(v)} a_{u\to v}(r_v^{(t)}-r_u^{(t)}), \quad r^{(t)}_v = \text{MLP}( h^{(t)}_v \oplus p_v^{(t)})
\] 
\[
a_{u\to v}=\quad\frac{\exp(s (r_v^{(t)},r_u^{(t)}))}{\sum_{u \in G_2(v)}\exp(s 
(r_v^{(t)},r_{u}^{(t)}))}, \quad s (r_v,r_u) = \frac{r_v \cdot r_u}{\sqrt{d}}
\] where $a_{u\to v}$ is the attention weight of node $v$ in $G_1$ to $u$ in $G_2$, $s$ is the similarity function, and $d$ is the dimension of $h$. As dimension \(d\) increases, leading to higher attention score variance, the softmax function enters a low-gradient region, impeding back-propagation learning \citep{vaswani2017attention}. The differences between graphs are encapsulated within the cross-graph matching vectors and will be accentuated throughout the propagation process.

\section{Experiment}
% in the section, we answer the following questions,
% \begin{itemize}
%     \item Q1: the effectiveness of \ours.
%     \item Q2: Does \textit{RelNode} help the SQL functional correctness?
%     \item Q3: Is our functional evaluation correctness better than other evaluation metrics?
% \end{itemize}
% In this section, we introduce our datasets, baselines, ablation studies and implementation details. The correlation evaluation metrics are discussed in Appendix \ref{a: correlation evaluation}.

\subsection{Dataset}
Based on the SQL-generated benchmark dataset Spider \citep{yu2018spider}, we constructed a new dataset, \texttt{Spider-Pair}, for code evaluation purposes. This dataset consists of a training set and two testing sets, each containing pairs of SQLs to simulate diverse scenarios for SQL code evaluation. Each pair is composed of the golden SQL and generated SQL as well as a ground truth label indicating the correctness of the generated SQL. The training set (train) and one testing dataset (test) is composed of data from NL2SQL code generation using large language models, while the other (test-aug) is generated from SQL equivalence rewriting. To mitigate the issue of data leakage within the evaluation dataset, we ensure that the training and testing sets originate from distinct databases. The more details are shown in Appendix \ref{sec: parispider}.

\subsection{Ablation Studies}

We evaluate code evaluation metrics against the code's functional correctness using correlation evaluation metrics such as \textbf{AUC}, \textbf{Spearman R} ($r_s$) and \textbf{Pearson R} ($r_p$) \cite{dong2023codescore}, as discussed in Appendix \ref{a: correlation evaluation}. In this section, we conduct ablation studies across four key areas to assess the impact of various components within our model. 
% These areas include comparing \textit{RelNode} to AST, evaluating the effectiveness of adding logic and data flows, examining various approaches for calculating positional encodings, and assessing the performance of different Graph Matching Neural Networks (GMNNs). The results are shown in Table \ref{ablation studies}.

\renewcommand{\arraystretch}{1.2} % Increase line spacing
\begin{table} [H]
\begin{adjustwidth}{0cm}{0cm} % 使用adjustwidth环境调整宽度
\centering
\small
\begin{tabular}{llllllll}
\toprule
\multirow{2}{*}{\textbf{Type}} &  \multirow{2}{*}{\textbf{Method}} & \multicolumn{3}{c}{\textbf{test}} & \multicolumn{3}{c}{\textbf{test-aug}} \\ \cline{3-8} 
&  &  AUC   & \(\tau\)   & \(r_s\)    & AUC    & \(\tau\)     & \(r_s\)   \\ 
\midrule
\multirow{5}{*}{\textbf{GMN}} & (0): AST  & 0.9053 & 0.7000 & 0.5718 & 0.5788 & 0.1365 & 0.1114 \\

& (1): \textit{RelNode} & 0.9201 & 0.7256 & 0.5926 & 0.7975 & 0.5153 & 0.4208 \\
& (2): (1) + logic + data & 0.9410 & 0.7617 & 0.6221 & 0.8154 & 0.5462 & 0.4461 \\
& (3): (2) + separated PE & 0.9326 & 0.7472 & 0.6103 & 0.8284 & 0.5689 & 0.4646 \\
& (4): (2) + global PE & \textbf{0.9432} & \textbf{0.7675} & \textbf{0.6352} & \textbf{0.8480} & \textbf{0.6033} & \textbf{0.4962} \\
\hdashline
\multirow{3}{*}{\textbf{{Other GMNNs}}} & (5): (2) + MGMN         & 0.8566   & 0.6159         & 0.5032            & 0.7360      & 0.4088           & 0.3338             \\
& (6): (2) + EGSC          & 0.8822   & 0.6600         & 0.5316            & 0.7656      & 0.4600           & 0.3758             \\
& (7): (2) + ERIC          & 0.9166   & 0.7202         & 0.5922            & 0.8051      & 0.5286           & 0.4329             \\
\bottomrule
\end{tabular}
\end{adjustwidth}

\caption{Ablation Studies. \textnormal{Experiment (0) parses SQL into an AST, whereas (1) parses SQL into a \textit{RelNode}; (2) incorporates control flow and data flow into the \textit{RelNode} to capture more semantic information. Experiments for (0), (1), and (2) are conducted on the original version of GMN; (3) and (4) build upon (2), and introduce different Positional Embeddings (PE) into the GMN's cross attention. The PE in (3) is calculated on separate graphs, while the PE in (4) is derived from the merged graph obtained by connecting seed nodes. The line charts in the Appendix \ref{auc: appendix} show the changes in the AUC of the test sets during the training of experiments (1)-(4).}}
\label{ablation studies}
\end{table}

Firstly, we use the mo\_sql\_parsing \footnote{\url{https://github.com/klahnakoski/mo-sql-parsing}} library to transform SQL into Abstract Syntax Trees (AST). In the test dataset, the generated and reference SQL codes tend to have similar syntactic structures, thus achieving an AUC of 90.53\%. 
However, on the enhanced test-aug dataset, the AUC drops significantly to 57.88\%. 
% AST only focuses on the syntactic and structural information, overlooking the semantic information.
To overcome the limitation of AST, we introduced \textit{RelNode}, a SQL optimization tool capable of identifying similar logical execution patterns across varying syntaxes. 
\textit{RelNode} improved performance by 39\% on the functional augmented test dataset. Additionally, integrating logic and data flow analysis into \textit{RelNode} allows us to uncover and convey deeper semantics behind the textual features, boosting performance by 2.2\% across both datasets.

Further enhancements to the Graph Matching Network (GMN) include learnable Positional Embeddings (PE) to include structural and positional information. Experiment (3), applying PE to two separate graphs, increased the AUC by 1.6\% on test-aug but decreased by 0.89\% on the test dataset, which indicates that emphasizing structural differences in SQL pairs with similar syntactic structures will lead to overfitting. 
In contrast, Experiment (4) calculates PE on a merged graph, leading to gains of 0.23\% and 4.0\% on the test and test-aug datasets, respectively. PE on the merged graph not only highlights differences between nodes at different levels in \textit{RelNode} but also brings features of potentially matching subtrees closer, thereby enhancing the model's ability to generalize beyond the training dataset.

In addition, we experimented with various Graph Matching Neural Networks (GMNNs), but their performance was inferior to that of GMN.  We speculate that in SQL evaluation task, compared to MGNN \citep{ling2021multilevel}, our optimized GMN leverages global positional encoding to enhance its ability to capture equivalent structures. Furthermore, compared to EGSC \citep{qin2021slow} and ERIC \citep{zhuo2022efficient}, it can better focus on subtle differences between nodes through cross-attention.

% but their performance was inferior to that of GMN. 
% We attribute this to the following reasons: MGNN \citep{ling2021multilevel} overly focuses on dense and fine-grained similarity calculations, neglecting the global information of equivalent structures; EGSC \citep{qin2021slow}  designed an attention-based collaborative feature fusion network on multi-level GNN features, but completely overlooked cross-graph node-level interactions. 
% ERIC \citep{zhuo2022efficient} applied alignment regularization to the GNN encoder to impose node-to-graph correspondence constraints, simplifying interactions between nodes. 
% Due to the absence of a node-level matching module, it is challenging to capture the subtle differences in key structures within the graph representation.

\subsection{Comparison with Other Evaluation Metrics}

% Component Matching Accuracy \citep{yu2018spider}, Exact Matching Accuracy \citep{yu2018spider}, BLEU \citep{papineni2002bleu}, BERTScore \citep{zhang2019bertscore}, COMET \citep{rei2020comet}, 

To comprehensively evaluate our model, we compare it with Matching-based Metrics including CrystalBLEU \citep{eghbali2022crystalbleu}, CodeBLEU \citep{ren2020codebleu}, \texttt{RelPM}, ASTPM and pre-trained model-based methods including CodeScore \citep{dong2023codescore}, CodeBERTScore \citep{zhou2023codebertscore} on two datasets. In addition, to evaluate from the perspective of SQL equivalence, we compare our approach with the state-of-the-art SQL equivalence verifier, SPES \cite{zhou2022spes}, and a custom-designed prompt for assessing SQL equivalence based on functional and logical consistency using the GPT-4 model.

\begin{table}[H]
\centering
\small
\begin{adjustwidth}{0cm}{0cm} % 使用adjustwidth环境调整宽度
\begin{tabular}{llllllll}
\toprule
\multirow{2}{*}{\textbf{Type}} & \multirow{2}{*}{\textbf{Method}} & \multicolumn{3}{c}{\textbf{test}} & \multicolumn{3}{c}{\textbf{test-aug}} \\ \cline{3-8} 
&  & AUC   & $\tau$  & $r_s$    & AUC    &  $\tau$     & $r_s$    \\ 
\midrule

\multirow{4}{*}{\textbf{Matching-based}} & CrystalBLEU  & 0.6522   & 0.2629        & 0.2148            &0.5255      & 0.0441           & 0.0360            \\ 
 & CodeBLEU        & 0.6801   & 0.3119        & 0.2548            &0.5793      & 0.1374           & 0.1122             \\ 
                     & ASTPM         & 0.6878   & 0.3289        & 0.2824            &0.5651      & 0.1145           & 0.0981             \\ 
                      & \texttt{RelPM} (Ours)      & 0.7817   & 0.4870        & 0.4011            &0.6237      & 0.2144           & 0.1759             \\ 

\midrule

\multirow{2}{*}{\textbf{Pre-trained Models}}    
 & CodeBERTScore   & 0.7259   & 0.3191         & 0.3907            & 0.5482      & 0.0672   & 0.0822             \\
 & CodeScore   & 0.8848   & 0.6665         & 0.5445            & 0.8351      & 0.5792           & 0.4731             \\  \midrule
{\textbf{Equivalence Verifier}}    
& SPES      & 0.7270   & 0.5052      & 0.5052           & 0.6585      & 0.3640           & 0.3640            \\  
& GPT-4      & 0.7960 & 0.5142 & 0.5577 & 0.7566 & 0.4341 & 0.4720 \\ \midrule                    
{\textbf{GNN}}    
& \texttt{FuncEvalGMN} (Ours)         & \textbf{0.9432}   & \textbf{0.7675}         & \textbf{0.6352}            & \textbf{0.8480}      & \textbf{0.6033}           & \textbf{0.4962}             \\ 

\bottomrule
\end{tabular}

\end{adjustwidth}
\caption{Comparative Analysis of Other Evaluation Metrics. }
\label{comparedev}
\end{table}

As Table \ref{comparedev} demonstrates, our proposed \texttt{RelPM} significantly surpasses all matching-based methods, with \texttt{FuncEvalGMN} achieving state-of-the-art (SOTA) levels. Compared to \texttt{RelPM}, ASTPM's performance drops by about 15\%, proving that \textit{RelNode} can mitigate the impact of syntactic differences and capture deeper semantic information under logical execution. To better exploit the potential of CodeBLEU, we replaced AST with \textit{RelNode}. However, CodeBLEU only performs coarse-grained subtree matching on syntax trees, and its performance is merely on par with ASTPM, highlighting the superiority of our proposed partial matching algorithm. Additionally, since CrystalBLEU disregards the syntactic structure and semantic information of code, it performs the worst among all matching-based algorithms.

Within the Pre-trained model-based category, CodeBERTScore relies solely on the prior knowledge embedded in the pre-trained model, resulting in poor scalability on new languages, such as SQL. Specifically, it achieves only a 54.82\% AUC on the test-aug dataset. In contrast, CodeScore fine-tunes the pre-trained model through data-driven learning, thereby achieving the best results on both datasets, except for our method. However, CodeScore treats code merely as a sequence of tokens, overlooking the unique syntactic structure and semantic information of the code. Moreover, it only performs a simple linear concatenation of the generated code with the reference code, hindering the perception of structural differences between code pairs. Nevertheless, due to the model's inherent prior knowledge, it performs impressively on the test-aug dataset, which has a different data distribution from the training set, scoring only 1.5\% lower than our 
\texttt{FuncEvalGMN}.

SPES involves generating symbolic representations of queries and establishing their equivalence by assessing query containment relationships among these symbolic representations. However, it doesn't fully support all SQL keywords and struggles to differentiate between case-sensitive strings, resulting in AUCs of only 72.7\% and 65.85\% on two test datasets. The performance of GPT-4 surpasses that of the SPES method, indicating that GPT-4 has a certain advantage in determining SQL equivalence. However, due to the inherent limitations in GPT-4's understanding of SQL statements, its effectiveness does not match the functional consistency evaluation provided by our proposed FuncEvalGMN method.

\subsection{Generalization to Other NL2SQL Dataset}

BIRD \cite{li2024can} is a Dataset which has more complex schemas which is more suitable for the LLM SQL generation task. We simulate the BIRD-pair dataset with the same guidelines in Appendix \ref{sec: data SQL llm}. 

The performance of FuncEvalGMN Table~\ref{tab:performance_comparison} remains superior compared to other methods, and hence generalizibility is evident. 
 Also, we evaluate complexities of two data sets, Spider-dev and BIRD-dev, and demonstrate their difference in distribution over keywords in Table~\ref{tab:keyword_performance} of Appendix~\ref{sec:keyword_distribution} . BIRD dev dataset has a significantly higher usage of JOIN and WHERE which shows a much more complex table structure and much more complex syntax structure in BIRD-dev dataset. Not surprisingly, there is a noticeable decline in the correlation metric on BIRD-dev compared to on Spider-dev due to this domain gap. Nevertheless, our two datasets, Spider-pair and BIRD-pair, form a basis for fine-tuning datasets and facilitate additional fine-tuning across different domains. Due to the correlation metric , our method is hopeful to serve as a general model for evaluation of Text2SQL generation. 

\begin{table}[h]
    \centering
    \begin{tabular}{llccc}
        \toprule
        Type                 & Method              & AUC    & $\tau$  & $r_s$  \\
        \midrule
        {\textbf{Matching-based}}       & CrystalBLEU         & 0.6368 & 0.1921 & 0.2352  \\
                             & CodeBLEU (with ROT) & 0.7418 & 0.3427 & 0.4162  \\
                             & ASTPM               & 0.6872 & 0.2662 & 0.3219  \\
                             & \texttt{RelPM} (Ours)              & 0.7186 & 0.3101 & 0.3763  \\
        \midrule
        {\textbf{Pre-trained Models}}   & CodeBertScore       & 0.7066 & 0.2900 & 0.3552  \\
                             & CodeScore           & 0.7920  & 0.3830  & 0.4330   \\ 
        \midrule
        {\textbf{Equivalence Verifier}} &
        GPT-4  & 0.8108 & 0.5054 & 0.5593 \\
        \midrule
        {\textbf{GNN}}       & 
        FuncEvalGMN(Ours)        & \textbf{0.8949} & \textbf{0.6078} & \textbf{0.6934}  \\
        \bottomrule
    \end{tabular}
    \caption{Performance Comparison of Different Methods on BIRD Dataset.}
    \label{tab:performance_comparison}
\end{table}

\section{Conclusion}
In our work, we present \texttt{FuncEvalGMN}, a novel graph-based approach for enhancing the evaluation of functional correctness in code, particularly SQL. This method involves transforming SQL into a graph that captures its syntactic and semantic essence, using a combination of node types and encoding strategies to represent SQL structures and parameters. We leverage Graph Matching Networks to assess graph similarity, incorporating positional embedding for improving structural understanding. Our experiments include a unique dataset, \texttt{Spider-Pair}, and baseline methods for comprehensive evaluation. Looking forward, we aim to extend this methodology to additional programming languages, exploring its potential to generalize across diverse coding paradigms and enhance code evaluation techniques further.

\newpage

\bibliography{colm2024_conference}

\begin{thebibliography}{57}
\providecommand{\natexlab}[1]{#1}
\providecommand{\url}[1]{\texttt{#1}}
\expandafter\ifx\csname urlstyle\endcsname\relax
  \providecommand{\doi}[1]{doi: #1}\else
  \providecommand{\doi}{doi: \begingroup \urlstyle{rm}\Url}\fi

\bibitem[Allamanis et~al.(2018)Allamanis, Brockschmidt, and
  Khademi]{allamanis2018learning}
Miltiadis Allamanis, Marc Brockschmidt, and Mahmoud Khademi.
\newblock Learning to represent programs with graphs.
\newblock In \emph{International Conference on Learning Representations}, 2018.

\bibitem[Bai et~al.(2019)Bai, Ding, Bian, Chen, Sun, and Wang]{bai2019simgnn}
Yunsheng Bai, Hao Ding, Song Bian, Ting Chen, Yizhou Sun, and Wei Wang.
\newblock Simgnn: A neural network approach to fast graph similarity
  computation.
\newblock In \emph{Proceedings of the twelfth ACM international conference on
  web search and data mining}, pp.\  384--392, 2019.

\bibitem[Bai et~al.(2020)Bai, Ding, Gu, Sun, and Wang]{bai2020learning}
Yunsheng Bai, Hao Ding, Ken Gu, Yizhou Sun, and Wei Wang.
\newblock Learning-based efficient graph similarity computation via multi-scale
  convolutional set matching.
\newblock In \emph{Proceedings of the AAAI Conference on Artificial
  Intelligence}, volume~34, pp.\  3219--3226, 2020.

\bibitem[Begoli et~al.(2018)Begoli, Camacho-Rodr{\'\i}guez, Hyde, Mior, and
  Lemire]{begoli2018apache}
Edmon Begoli, Jes{\'u}s Camacho-Rodr{\'\i}guez, Julian Hyde, Michael~J Mior,
  and Daniel Lemire.
\newblock Apache calcite: A foundational framework for optimized query
  processing over heterogeneous data sources.
\newblock In \emph{Proceedings of the 2018 International Conference on
  Management of Data}, pp.\  221--230, 2018.

\bibitem[Bojanowski et~al.(2017)Bojanowski, Grave, Joulin, and
  Mikolov]{bojanowski2017enriching}
Piotr Bojanowski, Edouard Grave, Armand Joulin, and Tomas Mikolov.
\newblock Enriching word vectors with subword information.
\newblock \emph{Transactions of the association for computational linguistics},
  5:\penalty0 135--146, 2017.

\bibitem[Cao et~al.(2023{\natexlab{a}})Cao, Chen, Li, Zhang, Xu, Zhang, and
  Yu]{cao2023heterogeneous}
Ruisheng Cao, Lu~Chen, Jieyu Li, Hanchong Zhang, Hongshen Xu, Wangyou Zhang,
  and Kai Yu.
\newblock A heterogeneous graph to abstract syntax tree framework for
  text-to-sql.
\newblock \emph{IEEE Transactions on Pattern Analysis and Machine
  Intelligence}, 2023{\natexlab{a}}.

\bibitem[Cao et~al.(2023{\natexlab{b}})Cao, Zhang, Xu, Li, Ma, Chen, and
  Yu]{cao2023astormer}
Ruisheng Cao, Hanchong Zhang, Hongshen Xu, Jieyu Li, Da~Ma, Lu~Chen, and Kai
  Yu.
\newblock Astormer: An ast structure-aware transformer decoder for text-to-sql.
\newblock \emph{arXiv preprint arXiv:2310.18662}, 2023{\natexlab{b}}.

\bibitem[Chen et~al.(2021{\natexlab{a}})Chen, Luo, Zhang, Zhou, Bai, Hu, Tai,
  and Quan]{chen2021learning}
Hongkai Chen, Zixin Luo, Jiahui Zhang, Lei Zhou, Xuyang Bai, Zeyu Hu, Chiew-Lan
  Tai, and Long Quan.
\newblock Learning to match features with seeded graph matching network.
\newblock In \emph{Proceedings of the IEEE/CVF International Conference on
  Computer Vision}, pp.\  6301--6310, 2021{\natexlab{a}}.

\bibitem[Chen et~al.(2021{\natexlab{b}})Chen, Tworek, Jun, Yuan, Pinto, Kaplan,
  Edwards, Burda, Joseph, Brockman, et~al.]{chen2021evaluating}
Mark Chen, Jerry Tworek, Heewoo Jun, Qiming Yuan, Henrique Ponde de~Oliveira
  Pinto, Jared Kaplan, Harri Edwards, Yuri Burda, Nicholas Joseph, Greg
  Brockman, et~al.
\newblock Evaluating large language models trained on code.
\newblock \emph{arXiv preprint arXiv:2107.03374}, 2021{\natexlab{b}}.

\bibitem[Chu et~al.(2018)Chu, Murphy, Roesch, Cheung, and
  Suciu]{chu2018axiomatic}
Shumo Chu, Brendan Murphy, Jared Roesch, Alvin Cheung, and Dan Suciu.
\newblock Axiomatic foundations and algorithms for deciding semantic
  equivalences of sql queries, 2018.

\bibitem[Cota et~al.(1994)Cota, Fritz, and Sargent]{cota1994control}
Bruce~A Cota, Douglas~G Fritz, and Robert~G Sargent.
\newblock Control flow graphs as a representation language.
\newblock In \emph{Proceedings of Winter Simulation Conference}, pp.\
  555--559. IEEE, 1994.

\bibitem[Cyganiak(2005)]{cyganiak2005relational}
Richard Cyganiak.
\newblock A relational algebra for sparql.
\newblock \emph{Digital Media Systems Laboratory HP Laboratories Bristol.
  HPL-2005-170}, 35\penalty0 (9), 2005.

\bibitem[Deng et~al.(2021)Deng, Awadallah, Meek, Polozov, Sun, and
  Richardson]{Deng_2021}
Xiang Deng, Ahmed~Hassan Awadallah, Christopher Meek, Oleksandr Polozov, Huan
  Sun, and Matthew Richardson.
\newblock Structure-grounded pretraining for text-to-sql.
\newblock In \emph{Proceedings of the 2021 Conference of the North American
  Chapter of the Association for Computational Linguistics: Human Language
  Technologies}. Association for Computational Linguistics, 2021.
\newblock \doi{10.18653/v1/2021.naacl-main.105}.
\newblock URL \url{http://dx.doi.org/10.18653/v1/2021.naacl-main.105}.

\bibitem[Dong et~al.(2023)Dong, Ding, Jiang, Li, Li, and
  Jin]{dong2023codescore}
Yihong Dong, Jiazheng Ding, Xue Jiang, Zhuo Li, Ge~Li, and Zhi Jin.
\newblock Codescore: Evaluating code generation by learning code execution.
\newblock \emph{arXiv preprint arXiv:2301.09043}, 2023.

\bibitem[Dwivedi et~al.(2021)Dwivedi, Luu, Laurent, Bengio, and
  Bresson]{dwivedi2021graph}
Vijay~Prakash Dwivedi, Anh~Tuan Luu, Thomas Laurent, Yoshua Bengio, and Xavier
  Bresson.
\newblock Graph neural networks with learnable structural and positional
  representations.
\newblock \emph{arXiv preprint arXiv:2110.07875}, 2021.

\bibitem[Eghbali \& Pradel(2022)Eghbali and Pradel]{eghbali2022crystalbleu}
Aryaz Eghbali and Michael Pradel.
\newblock Crystalbleu: precisely and efficiently measuring the similarity of
  code.
\newblock In \emph{Proceedings of the 37th IEEE/ACM International Conference on
  Automated Software Engineering}, pp.\  1--12, 2022.

\bibitem[Fang et~al.(2020)Fang, Liu, Shi, Huang, and Shi]{fang2020functional}
Chunrong Fang, Zixi Liu, Yangyang Shi, Jeff Huang, and Qingkai Shi.
\newblock Functional code clone detection with syntax and semantics fusion
  learning.
\newblock In \emph{Proceedings of the 29th ACM SIGSOFT international symposium
  on software testing and analysis}, pp.\  516--527, 2020.

\bibitem[Feng et~al.(2020)Feng, Guo, Tang, Duan, Feng, Gong, Shou, Qin, Liu,
  Jiang, et~al.]{feng2020codebert}
Zhangyin Feng, Daya Guo, Duyu Tang, Nan Duan, Xiaocheng Feng, Ming Gong, Linjun
  Shou, Bing Qin, Ting Liu, Daxin Jiang, et~al.
\newblock Codebert: A pre-trained model for programming and natural languages.
\newblock \emph{arXiv preprint arXiv:2002.08155}, 2020.

\bibitem[Finegan-Dollak et~al.(2018)Finegan-Dollak, Kummerfeld, Zhang,
  Ramanathan, Sadasivam, Zhang, and Radev]{Finegan_Dollak_2018}
Catherine Finegan-Dollak, Jonathan~K. Kummerfeld, Li~Zhang, Karthik Ramanathan,
  Sesh Sadasivam, Rui Zhang, and Dragomir Radev.
\newblock Improving text-to-sql evaluation methodology.
\newblock In \emph{Proceedings of the 56th Annual Meeting of the Association
  for Computational Linguistics (Volume 1: Long Papers)}. Association for
  Computational Linguistics, 2018.
\newblock \doi{10.18653/v1/p18-1033}.
\newblock URL \url{http://dx.doi.org/10.18653/v1/P18-1033}.

\bibitem[He et~al.(2016)He, Zhang, Ren, and Sun]{he2016deep}
Kaiming He, Xiangyu Zhang, Shaoqing Ren, and Jian Sun.
\newblock Deep residual learning for image recognition.
\newblock In \emph{Proceedings of the IEEE conference on computer vision and
  pattern recognition}, pp.\  770--778, 2016.

\bibitem[Huang \& Ling(2005)Huang and Ling]{huang2005using}
Jin Huang and Charles~X Ling.
\newblock Using auc and accuracy in evaluating learning algorithms.
\newblock \emph{IEEE Transactions on knowledge and Data Engineering},
  17\penalty0 (3):\penalty0 299--310, 2005.

\bibitem[Iyer et~al.(2017)Iyer, Konstas, Cheung, Krishnamurthy, and
  Zettlemoyer]{iyer2017learning}
Srinivasan Iyer, Ioannis Konstas, Alvin Cheung, Jayant Krishnamurthy, and Luke
  Zettlemoyer.
\newblock Learning a neural semantic parser from user feedback, 2017.

\bibitem[Kendall(1938)]{kendall1938new}
Maurice~G Kendall.
\newblock A new measure of rank correlation.
\newblock \emph{Biometrika}, 30\penalty0 (1/2):\penalty0 81--93, 1938.

\bibitem[Kulal et~al.(2019)Kulal, Pasupat, Chandra, Lee, Padon, Aiken, and
  Liang]{kulal2019spoc}
Sumith Kulal, Panupong Pasupat, Kartik Chandra, Mina Lee, Oded Padon, Alex
  Aiken, and Percy~S Liang.
\newblock Spoc: Search-based pseudocode to code.
\newblock \emph{Advances in Neural Information Processing Systems}, 32, 2019.

\bibitem[Li et~al.(2024)Li, Hui, Qu, Yang, Li, Li, Wang, Qin, Geng, Huo,
  et~al.]{li2024can}
Jinyang Li, Binyuan Hui, Ge~Qu, Jiaxi Yang, Binhua Li, Bowen Li, Bailin Wang,
  Bowen Qin, Ruiying Geng, Nan Huo, et~al.
\newblock Can llm already serve as a database interface? a big bench for
  large-scale database grounded text-to-sqls.
\newblock \emph{Advances in Neural Information Processing Systems}, 36, 2024.

\bibitem[Li et~al.(2015)Li, Tarlow, Brockschmidt, and Zemel]{li2015gated}
Yujia Li, Daniel Tarlow, Marc Brockschmidt, and Richard Zemel.
\newblock Gated graph sequence neural networks.
\newblock \emph{arXiv preprint arXiv:1511.05493}, 2015.

\bibitem[Li et~al.(2019)Li, Gu, Dullien, Vinyals, and Kohli]{li2019graph}
Yujia Li, Chenjie Gu, Thomas Dullien, Oriol Vinyals, and Pushmeet Kohli.
\newblock Graph matching networks for learning the similarity of graph
  structured objects.
\newblock In \emph{International conference on machine learning}, pp.\
  3835--3845. PMLR, 2019.

\bibitem[Ling et~al.(2021)Ling, Wu, Wang, Ma, Xu, Liu, Wu, and
  Ji]{ling2021multilevel}
Xiang Ling, Lingfei Wu, Saizhuo Wang, Tengfei Ma, Fangli Xu, Alex~X Liu,
  Chunming Wu, and Shouling Ji.
\newblock Multilevel graph matching networks for deep graph similarity
  learning.
\newblock \emph{IEEE Transactions on Neural Networks and Learning Systems},
  2021.

\bibitem[Liu et~al.(2020)Liu, Mao, Zhang, Xie, Wang, and Zhang]{liu2020graph}
Chunxiao Liu, Zhendong Mao, Tianzhu Zhang, Hongtao Xie, Bin Wang, and Yongdong
  Zhang.
\newblock Graph structured network for image-text matching.
\newblock In \emph{Proceedings of the IEEE/CVF conference on computer vision
  and pattern recognition}, pp.\  10921--10930, 2020.

\bibitem[Liu et~al.(2022)Liu, Li, Wei, Xia, Fu, and Jin]{liu2022unified}
Fang Liu, Ge~Li, Bolin Wei, Xin Xia, Zhiyi Fu, and Zhi Jin.
\newblock A unified multi-task learning model for ast-level and token-level
  code completion.
\newblock \emph{Empirical Software Engineering}, 27\penalty0 (4):\penalty0 91,
  2022.

\bibitem[Mi et~al.(2023)Mi, Zhan, Weng, Bao, Cui, and Ma]{mi2023graph}
Qing Mi, Yi~Zhan, Han Weng, Qinghang Bao, Longjie Cui, and Wei Ma.
\newblock A graph-based code representation method to improve code readability
  classification.
\newblock \emph{Empirical Software Engineering}, 28\penalty0 (4):\penalty0 87,
  2023.

\bibitem[Mikolov et~al.(2013)Mikolov, Chen, Corrado, and
  Dean]{mikolov2013efficient}
Tomas Mikolov, Kai Chen, Greg Corrado, and Jeffrey Dean.
\newblock Efficient estimation of word representations in vector space.
\newblock \emph{arXiv preprint arXiv:1301.3781}, 2013.

\bibitem[Orailoglu \& Gajski(1986)Orailoglu and
  Gajski]{orailoglu1986dataflowgraphrepresentation}
Alex Orailoglu and Daniel~D Gajski.
\newblock Flow graph representation.
\newblock In \emph{Proceedings of the 23rd ACM/IEEE Design Automation
  Conference}, pp.\  503--509, 1986.

\bibitem[Ottenstein \& Ottenstein(1984)Ottenstein and
  Ottenstein]{ottenstein1984programdependencegraph}
Karl~J Ottenstein and Linda~M Ottenstein.
\newblock The program dependence graph in a software development environment.
\newblock \emph{ACM Sigplan Notices}, 19\penalty0 (5):\penalty0 177--184, 1984.

\bibitem[Papineni et~al.(2002)Papineni, Roukos, Ward, and
  Zhu]{papineni2002bleu}
Kishore Papineni, Salim Roukos, Todd Ward, and Wei-Jing Zhu.
\newblock Bleu: a method for automatic evaluation of machine translation.
\newblock In \emph{Proceedings of the 40th annual meeting of the Association
  for Computational Linguistics}, pp.\  311--318, 2002.

\bibitem[Pranklin(1974)]{pranklin1974introduction}
A~Pranklin.
\newblock \emph{Introduction to the Theory of Statistics}.
\newblock 1974.

\bibitem[Qin et~al.(2021)Qin, Zhao, Wang, Wang, Zhang, and Fu]{qin2021slow}
Can Qin, Handong Zhao, Lichen Wang, Huan Wang, Yulun Zhang, and Yun Fu.
\newblock Slow learning and fast inference: Efficient graph similarity
  computation via knowledge distillation.
\newblock \emph{Advances in Neural Information Processing Systems},
  34:\penalty0 14110--14121, 2021.

\bibitem[Ren et~al.(2020)Ren, Guo, Lu, Zhou, Liu, Tang, Sundaresan, Zhou,
  Blanco, and Ma]{ren2020codebleu}
Shuo Ren, Daya Guo, Shuai Lu, Long Zhou, Shujie Liu, Duyu Tang, Neel
  Sundaresan, Ming Zhou, Ambrosio Blanco, and Shuai Ma.
\newblock Codebleu: a method for automatic evaluation of code synthesis.
\newblock \emph{arXiv preprint arXiv:2009.10297}, 2020.

\bibitem[Riba et~al.(2018)Riba, Fischer, Llad{\'o}s, and
  Forn{\'e}s]{riba2018learning}
Pau Riba, Andreas Fischer, Josep Llad{\'o}s, and Alicia Forn{\'e}s.
\newblock Learning graph distances with message passing neural networks.
\newblock In \emph{2018 24th International Conference on Pattern Recognition
  (ICPR)}, pp.\  2239--2244. IEEE, 2018.

\bibitem[Shi et~al.(2023)Shi, Cai, Zhao, Gao, Sood, and Xiang]{shi2023coss}
Chaochen Shi, Borui Cai, Yao Zhao, Longxiang Gao, Keshav Sood, and Yong Xiang.
\newblock Coss: leveraging statement semantics for code summarization.
\newblock \emph{IEEE Transactions on Software Engineering}, 2023.

\bibitem[Tang et~al.(2021)Tang, Li, Ge, Shen, Zhu, and Luo]{tang2021ast}
Ze~Tang, Chuanyi Li, Jidong Ge, Xiaoyu Shen, Zheling Zhu, and Bin Luo.
\newblock Ast-transformer: Encoding abstract syntax trees efficiently for code
  summarization.
\newblock In \emph{2021 36th IEEE/ACM International Conference on Automated
  Software Engineering (ASE)}, pp.\  1193--1195. IEEE, 2021.

\bibitem[Tang et~al.(2022)Tang, Shen, Li, Ge, Huang, Zhu, and Luo]{tang2022ast}
Ze~Tang, Xiaoyu Shen, Chuanyi Li, Jidong Ge, Liguo Huang, Zhelin Zhu, and Bin
  Luo.
\newblock Ast-trans: Code summarization with efficient tree-structured
  attention.
\newblock In \emph{Proceedings of the 44th International Conference on Software
  Engineering}, pp.\  150--162, 2022.

\bibitem[Vaswani et~al.(2017)Vaswani, Shazeer, Parmar, Uszkoreit, Jones, Gomez,
  Kaiser, and Polosukhin]{vaswani2017attention}
Ashish Vaswani, Noam Shazeer, Niki Parmar, Jakob Uszkoreit, Llion Jones,
  Aidan~N Gomez, {\L}ukasz Kaiser, and Illia Polosukhin.
\newblock Attention is all you need.
\newblock \emph{Advances in neural information processing systems}, 30, 2017.

\bibitem[Wang et~al.(2020)Wang, Li, Ma, Xia, and Jin]{wang2020detecting}
Wenhan Wang, Ge~Li, Bo~Ma, Xin Xia, and Zhi Jin.
\newblock Detecting code clones with graph neural network and flow-augmented
  abstract syntax tree.
\newblock In \emph{2020 IEEE 27th International Conference on Software
  Analysis, Evolution and Reengineering (SANER)}, pp.\  261--271. IEEE, 2020.

\bibitem[Wang \& Li(2021)Wang and Li]{wang2021code}
Yanlin Wang and Hui Li.
\newblock Code completion by modeling flattened abstract syntax trees as
  graphs.
\newblock In \emph{Proceedings of the AAAI conference on artificial
  intelligence}, volume~35, pp.\  14015--14023, 2021.

\bibitem[Wang et~al.(2022)Wang, Zhou, Yang, Ding, Hu, Ding, Tang, Chen, and
  Li]{10.1145/3514221.3526125}
Zhaoguo Wang, Zhou Zhou, Yicun Yang, Haoran Ding, Gansen Hu, Ding Ding, Chuzhe
  Tang, Haibo Chen, and Jinyang Li.
\newblock Wetune: Automatic discovery and verification of query rewrite rules.
\newblock In \emph{Proceedings of the 2022 International Conference on
  Management of Data}, SIGMOD '22, pp.\  94–107, New York, NY, USA, 2022.
  Association for Computing Machinery.
\newblock ISBN 9781450392495.
\newblock \doi{10.1145/3514221.3526125}.
\newblock URL \url{https://doi.org/10.1145/3514221.3526125}.

\bibitem[Xu et~al.(2022)Xu, Alon, Neubig, and Hellendoorn]{xu2022systematic}
Frank~F Xu, Uri Alon, Graham Neubig, and Vincent~Josua Hellendoorn.
\newblock A systematic evaluation of large language models of code.
\newblock In \emph{Proceedings of the 6th ACM SIGPLAN International Symposium
  on Machine Programming}, pp.\  1--10, 2022.

\bibitem[Xu et~al.(2017)Xu, Liu, Feng, Yin, Song, and Song]{xu2017neural}
Xiaojun Xu, Chang Liu, Qian Feng, Heng Yin, Le~Song, and Dawn Song.
\newblock Neural network-based graph embedding for cross-platform binary code
  similarity detection.
\newblock In \emph{Proceedings of the 2017 ACM SIGSAC conference on computer
  and communications security}, pp.\  363--376, 2017.

\bibitem[Yaghmazadeh et~al.(2017)Yaghmazadeh, Wang, Dillig, and
  Dillig]{yaghmazadeh2017sqlizer}
Navid Yaghmazadeh, Yuepeng Wang, Isil Dillig, and Thomas Dillig.
\newblock Sqlizer: query synthesis from natural language.
\newblock \emph{Proceedings of the ACM on Programming Languages}, 1\penalty0
  (OOPSLA):\penalty0 1--26, 2017.

\bibitem[Yu et~al.(2023)Yu, Yang, Chen, Chen, and Xu]{yu2023graph}
Dongjin Yu, Quanxin Yang, Xin Chen, Jie Chen, and Yihang Xu.
\newblock Graph-based code semantics learning for efficient semantic code clone
  detection.
\newblock \emph{Information and Software Technology}, 156:\penalty0 107130,
  2023.

\bibitem[Yu et~al.(2018)Yu, Zhang, Yang, Yasunaga, Wang, Li, Ma, Li, Yao,
  Roman, et~al.]{yu2018spider}
Tao Yu, Rui Zhang, Kai Yang, Michihiro Yasunaga, Dongxu Wang, Zifan Li, James
  Ma, Irene Li, Qingning Yao, Shanelle Roman, et~al.
\newblock Spider: A large-scale human-labeled dataset for complex and
  cross-domain semantic parsing and text-to-sql task.
\newblock \emph{arXiv preprint arXiv:1809.08887}, 2018.

\bibitem[Yu et~al.(2020)Yu, Zheng, Wang, Tang, Nie, and Wu]{yu2020codecmr}
Zeping Yu, Wenxin Zheng, Jiaqi Wang, Qiyi Tang, Sen Nie, and Shi Wu.
\newblock Codecmr: Cross-modal retrieval for function-level binary source code
  matching.
\newblock \emph{Advances in Neural Information Processing Systems},
  33:\penalty0 3872--3883, 2020.

\bibitem[Zhong et~al.(2020)Zhong, Yu, and Klein]{zhong2020semantic}
Ruiqi Zhong, Tao Yu, and Dan Klein.
\newblock Semantic evaluation for text-to-sql with distilled test suites.
\newblock \emph{arXiv preprint arXiv:2010.02840}, 2020.

\bibitem[Zhou et~al.(2022)Zhou, Arulraj, Navathe, Harris, and Wu]{zhou2022spes}
Qi~Zhou, Joy Arulraj, Shamkant~B Navathe, William Harris, and Jinpeng Wu.
\newblock Spes: A symbolic approach to proving query equivalence under bag
  semantics.
\newblock In \emph{2022 IEEE 38th International Conference on Data Engineering
  (ICDE)}, pp.\  2735--2748. IEEE, 2022.

\bibitem[Zhou et~al.(2023)Zhou, Alon, Agarwal, and
  Neubig]{zhou2023codebertscore}
Shuyan Zhou, Uri Alon, Sumit Agarwal, and Graham Neubig.
\newblock Codebertscore: Evaluating code generation with pretrained models of
  code.
\newblock \emph{arXiv preprint arXiv:2302.05527}, 2023.

\bibitem[Zhuo \& Tan(2022)Zhuo and Tan]{zhuo2022efficient}
Wei Zhuo and Guang Tan.
\newblock Efficient graph similarity computation with alignment regularization.
\newblock \emph{Advances in Neural Information Processing Systems},
  35:\penalty0 30181--30193, 2022.

\bibitem[Zhuo et~al.(2021)Zhuo, Cai, Zhang, and Lv]{zhuo2021long}
Zhongliu Zhuo, T~Cai, Xiaosong Zhang, and Fengmao Lv.
\newblock Long short-term memory on abstract syntax tree for sql injection
  detection.
\newblock \emph{IET Software}, 15\penalty0 (2):\penalty0 188--197, 2021.

\end{thebibliography}
\bibliographystyle{colm2024_conference}

\newpage
\appendix
\section{RelNode Partial Matching (\texttt{RelPM})}\label{arpm}

\texttt{RelPM} calculates the matched nodes and their scores for each \textit{RelNode}, using the F-beta score to determine the similarity between the two SQLs. We divide the process into three parts: Node Matching, RelNode Scoring, and Similarity Evaluation.

\subsection{Node Matching}

When matching two SQL queries' \textit{RelNodes}, we can designate one as the source tree and the other as the target tree, then use depth-first search to find their matching node sets. When evaluating the match for the target tree, the source tree is treated as a reference to measure how well the nodes of the target tree align with it. The success of node matching is determined by scoring against \( l \) candidate matching nodes, with the highest-scoring node being selected as the final match. The formula for this process is $ m = \max \{m^j\},  j \in \{0, 1, ... l\} $ where \( m^j \) represents matching score of node \( n \) with a candidate matching node \( n_j \) in another tree, which can be calculated as:
 
\[ m^j = \alpha \times m_{\text{self}}^j + (1 - \alpha) \frac{\sum_{i = 0}^{N} {m_{\text{child}^{i}}^j}}{N} \quad \alpha \in (0,1) \] 

\begin{figure}[H]
    \centering
    \includegraphics[width=1\linewidth, scale=1.00]{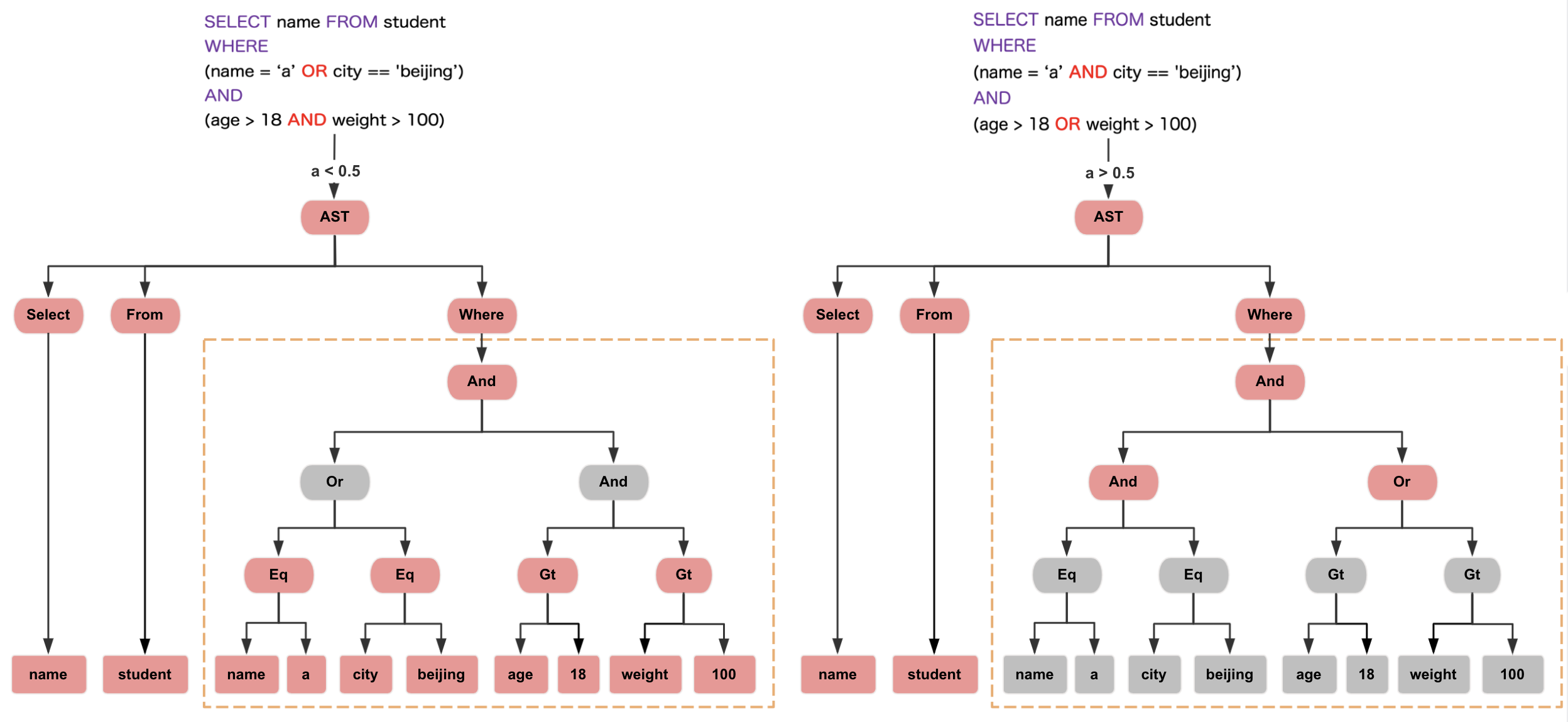}
    \caption{Partial Matching}
    \label{partial matching}
\end{figure}

The \( m_{\text{self}}^j \) represents the matching score of the node itself with the candidate matching node, and \( m_{\text{child}}^j \) indicates the maximum matching score of child node between two trees. The calculation of \( m_{\text{child}} \) is recursive, following the same method as \( m \), until it reaches leaf nodes. Additionally, as Figure \ref{partial matching} illustrates that \( \alpha \) serves as a weighting coefficient to balance the significance between the scores of a node and those of its children. A node colored in red signifies a successful match, whereas a gray node denotes an unsuccessful one. When \( \alpha \) falls below 0.5, the emphasis shifts towards the matching outcomes of the child nodes, leading us to circumvent the "OR" and "AND" logic in favor of matching a greater number of child nodes. Conversely, when \( \alpha \) exceeds 0.5, the focus is placed on matching the parent node, resulting in the failure to match for all corresponding subtree nodes beneath it.
\[
m_{\text{self}}^j = 
\begin{cases} 
1 & \text{if val}(n) = \text{val}(n_j) \\
0 & \text{otherwise}
\end{cases}
\]
For a given node, its matching score is determined by the candidate matching node \( n_j \). If the attributes of both nodes are identical, it is considered that a match can be established.

\subsection{RelNode Scoring}

For trees that exhibit matching information, we can adjust the weights of clauses and key nodes according to their importance, thereby influencing the overall score. The scoring equation is defined as:

\[ s = \omega \times s_{\text{self}} + \frac{\sum_{i = 0} ^ N \omega^{i} \times s_{\text{child}}^{i}}{N}, \quad \text{where} \quad \omega + \sum_{i = 0} ^ N \omega^{i} = 1 \]
\[ s_{\text{self}} = \sigma, \quad \sigma \in (0,1) \]

Here, \( \omega \) represents the weighting factor, and \( s_{\text{self}} \) denotes the node's own score, which is calculated during the process of matching stage. The calculation starts from the root node and proceeds recursively, combining the scores of the node itself and its children through a weighted sum.

\subsection{Similarity Evaluation}

By performing a cross-comparison between the source and target trees, 
"Precision" calculates the percentage of nodes in the source tree that successfully find matches in the target tree, while "recall" measures the percentage of nodes in the target tree that are matched in the source tree. %To assess similarity, we utilize \(F_{\beta}\), a similarity metric that represents the weighted harmonic mean of precision and recall, calculated as follows:
We compute the weighted geometric mean,

\[
F_{\beta} = \frac{(1 + \beta^2) \times \text{Precision} \times \text{Recall}}{\beta^2 \times \text{Precision} + \text{Recall}}
\]

When assessing code generation models, our main focus is on verifying if the generated code fully aligns with the semantics of the reference code. In this context, the reference code acts as the target tree, while the generated code represents the source tree. Hence, we give priority to recall in our evaluation, indicating that we assign a relatively high value to \( \beta \).

\section{Content Node Feature Embedding}
\begin{figure}[H]
    \centering
    \includegraphics[width=0.45\linewidth, scale=0.80]{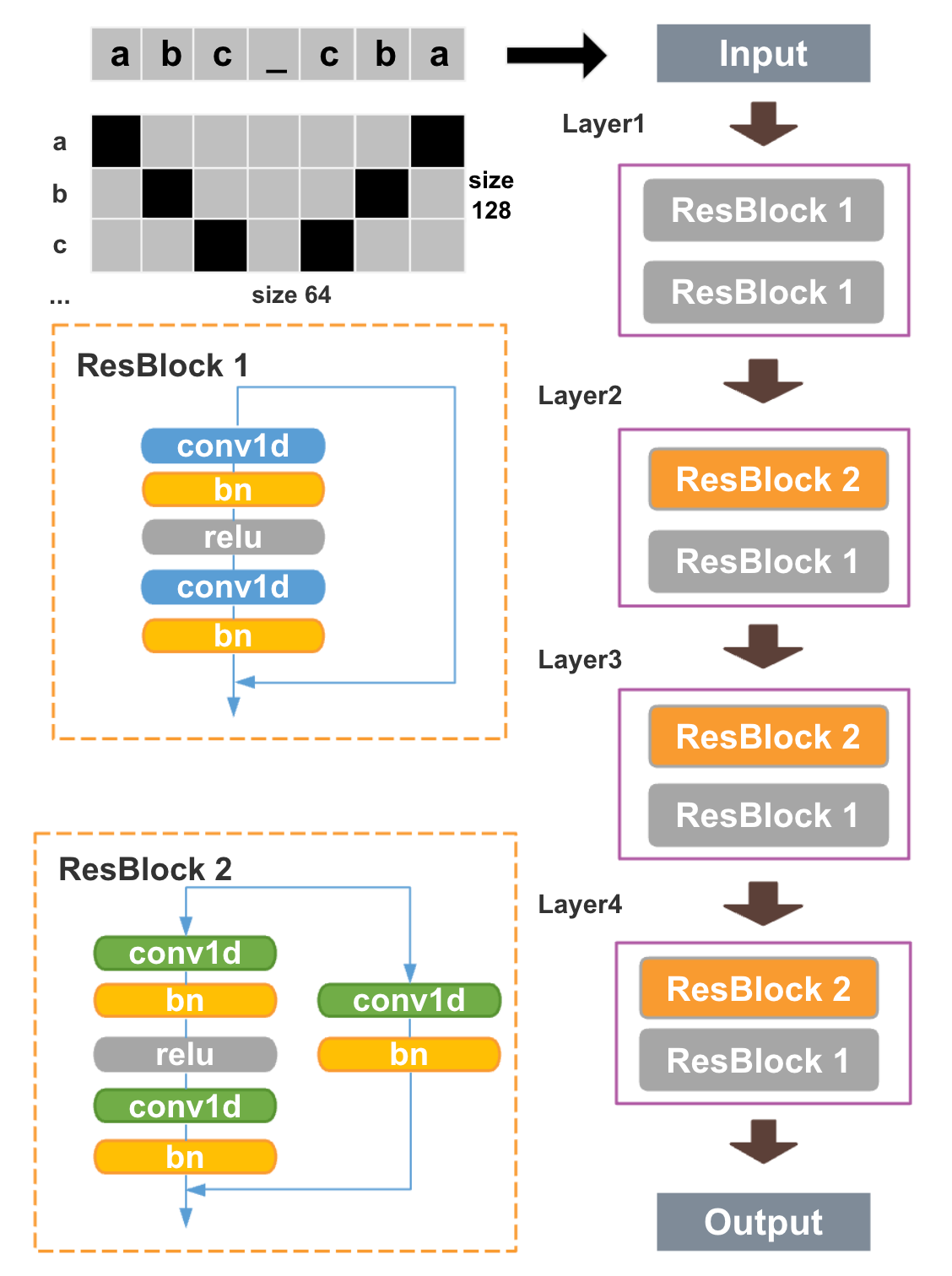}
    \caption{The Architecture for Content Node Feature Embedding}
    \label{resnet}
\end{figure}

\label{aresnet}

Figure \ref{resnet} depicts the process of representing a \textit{Content Node}, using "abc\_cba" as an example. Initially, the string is encoded into a one-dimensional vector through ASCII encoding. Following this, each element of the vector is transformed via one-hot encoding. As a result, "abc\_cba" is encoded into a 64 $\times$ 128 matrix. In this context, 64 denotes the string's length, with any deficiency up to 64 being compensated by zero padding. The 128 dimension reflects the ASCII one-hot encoding's dimensionality. To enhance the characterization of \textit{Content Nodes}, we used a ResNet \citep{he2016deep} model. This model comprises eight residual blocks (organized into 4 layers), with each block primarily employing 1D convolutional layers as its core components.

\section{RelNode Graph Matching Network}
\label{argmn}

In addition to message passing, the other implementation details of GMN are as follows: 
\subsection{Update Function} 

The update function integrates all gathered messages to update each node's representation at every iteration step. This process is mathematically formulated as:

\[
h^{(t+1)}_v = f_\mathrm{update}(h^{(t)}_v, m^{(t+1)}_v, \mu^{(t+1)}_v),
\]

In this equation, \(h^{(t+1)}_v\) is the updated representation of node \(v\) at step \(t+1\). The function \(f_\mathrm{update}\), which in our implementation is a Gated Recurrent Unit (GRU), updates the node's feature representation using its previous state \(h^{(t)}_v\), the inner-graph message \(m^{(t+1)}_v\), and cross-graph communication \(\mu^{(t+1)}_v\).

\subsection{Aggregator}
% Aggregation is to compute a graph-level representation. After a certain number of $T$ steps of propagation, an aggregator takes the set of node representations $\{h_v^{(T)}\}$ as input and computes a graph level representation $h_G= f_\mathrm{agg}(\{h_v^{(T)}\})$. We use the following aggregation method:
% \[
% h_G = MLP_G\left(\sum_{v\in G(v)} \sigma(MLP_\mathrm{gate}(h_v^{(T)})) \odot MLP(h_v^{(T)})\right),
% \]
% which transforms node representations and then uses a weighted sum with gating vectors to aggregate across nodes. The weighted sum can help filter out irrelevant information, it is more powerful than a simple sum and also works significantly better empirically.

Aggregation is to calculate a representation for the entire graph. Following \(T\) propagation steps, an aggregation function processes the set of node representations to produce a graph-level representation $h_G$. The aggregation method we utilize is proposed in \citep{li2015gated}:

\[
h_G = MLP_G\left(\sum_{v\in G(v)} \sigma(MLP_\mathrm{gate}(h_v^{(T)})) \odot MLP(h_v^{(T)})\right),
\]

In this method, a gated weighted sum, which aggregates information across all nodes and filters out irrelevant data, proves to be more effective than a simple summation.

\subsection{Similarity Metric}

Once we obtain the graph representations, \(h_{G_1}\) and \(h_{G_2}\), for the graph pair \((G_1, G_2)\), we evaluate their similarity using a vector space metric. Suitable metrics include Euclidean, cosine, or Hamming similarities. In our case, we employ the Euclidean distance as the similarity metric, which is defined as:

\[
s(h_{G_1}, h_{G_2}) = \| h_{G_1} - h_{G_2} \|_2.
\]

\subsection{Loss Function}
In our work, we need to evaluate the distance between the graph similarity we calculated and the true labels. To achieve this, we utilize a margin-based pairwise loss:

$$
L = \max\{0, \gamma - t(1 - s(h_{G_1}, h_{G_2}))\}
$$ where $t \in \{-1, 1\}$ denotes the label for the graph pair, and $\gamma > 0$ serves as a margin parameter. This loss function encourages similarity score $s(h_{G_1}, h_{G_2})$ is less than $1 - \gamma$ for similar pairs (when $t = 1$), and greater than $1 + \gamma$ for dissimilar pairs (when $t = -1$). This loss is robust to noisy data and doesn't over-penalize minor deviations. It also enhances model generalization by maintaining a margin, crucial for performance on unseen data, rather than merely minimizing training error.

\section{Equivalent Conversion Capabilities of RelNode}
\label{a:relnode}

% Calcite converts the relational algebraic expressions generated by the parser into an execution plan(RelNode structure) for the execution engine to execute, and in the process applies several rule optimizations to help generate more efficient execution plans.
% These optimization rules allow for the equivalent conversion of relational expressions. Common optimization rules include Predicate Pushdown, Constant Folding and Column Pruning.

Calcite converts relational algebra expressions generated by the parser into an execution plan, applying several optimization rules in the process. These optimization rules enable equivalent transformations of relational expressions, such as Predicate Pushdown, Constant Folding and Column Pruning.

\begin{figure}
    \centering
    \includegraphics[width=0.9\linewidth, scale=1.00]{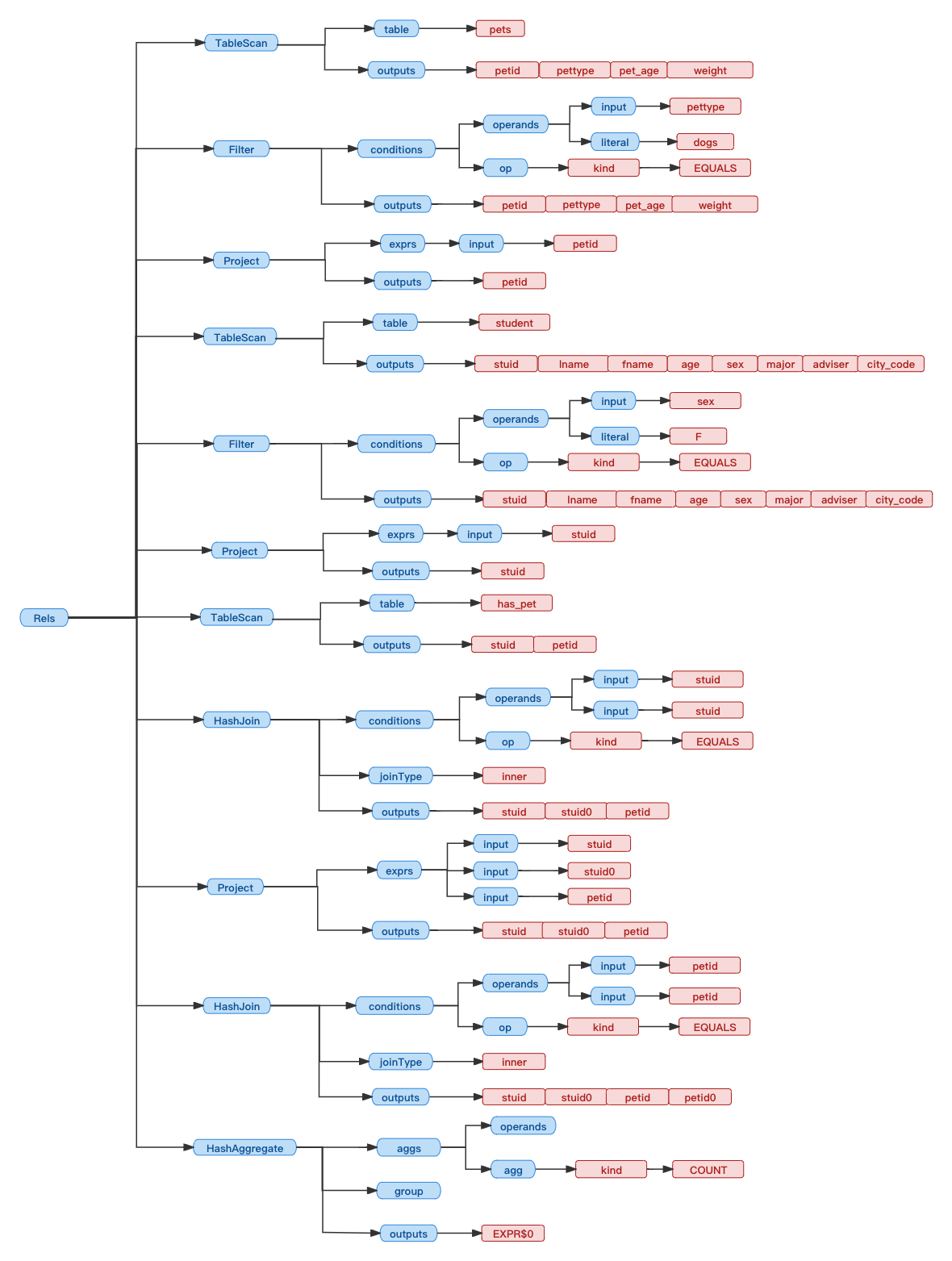}
    \caption{RelNode Structure of optimized SQL. }
    \label{relnode_eq}
\end{figure}

% In our work, the execution plan generated by Calcite is used for graph construction, which, on the one hand, can show more details of SQL execution; on the other hand, through the equivalence transformation function of Calcite, SQL can be standardized to a certain extent, which reduces the difficulty of the model to judge the functional equivalence.

In our work, we utilize execution plans generated by Calcite to construct graphs. It abstracts the syntactic structure of SQL and provides rich semantic information from the perspectives of logical execution and variable usage. Furthermore, its optimization of execution plans standardizes SQL, uncovering the same functionality under different syntactic structures, thereby reducing the difficulty of model judgment in determining functional equivalence.

\begin{figure}[H]
    \centering
    \includegraphics[width=\linewidth, scale=1.00]{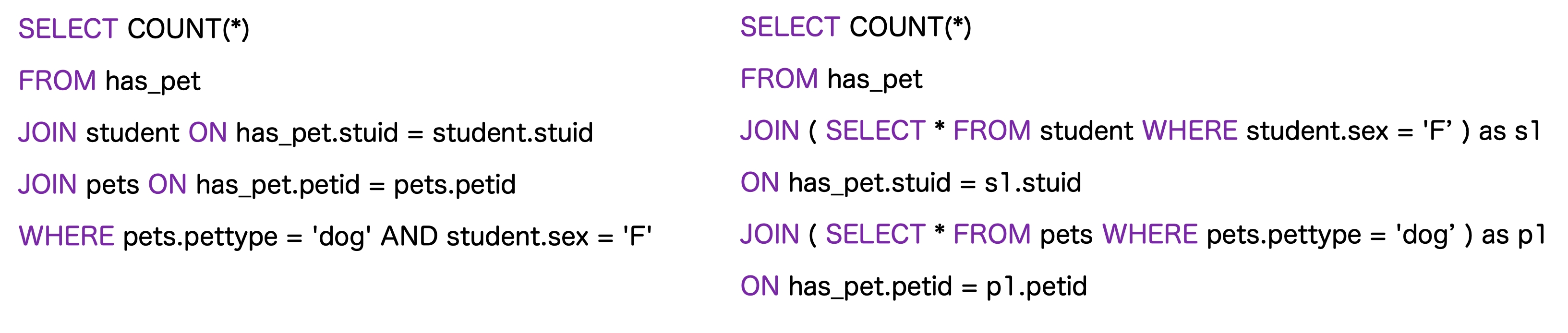}
    \caption{An Example of Predicate Pushdown }
    \label{relnode_sql}
\end{figure}

% Next, we will show a case of predicate pushdown optimization to demonstrate the equivalence transformation capability of \textit{RelNode}. Predicate pushdown refers to moving predicates in the \texttt{WHERE} clause of an outer query block into the lower-level query block in which it is contained, allowing for earlier data filtering and better use of indexes. Figure \ref{relnode_sql} shows two equivalent SQLs and figure \ref{relnode_eq} shows their corresponding \textit{RelNode} structures. The first SQL is to first perform $\theta$-join on the \texttt{has\_pet}, \texttt{student}, \texttt{pet} tables and then filter using the \texttt{pets.pettype = 'dog'} AND \texttt{student.sex = 'F'} condition. The predicate pushdown optimization will first apply the \texttt{pets.pettype = 'dog'} AND \texttt{student.sex = 'F'} condition to the \texttt{pet} and \texttt{student} tables for filtering before performing the $\theta$-join, the optimized SQL is shown in the second SQL.

In the following discussion, we explore a case of predicate pushdown optimization. Predicate pushdown is a strategy that involves relocating predicates from the \texttt{WHERE} clause of an outer query block to a more granular query block where the predicate is relevant. This approach enables earlier data filtering and enhances index utilization. Figures \ref{relnode_sql} demonstrate two SQL queries that are functionally equivalent, while Figure \ref{relnode_eq} illustrates their identical \textit{RelNode}.

The initial SQL query conducts a $\theta$-join across the \texttt{has\_pet}, \texttt{student}, and \texttt{pets} tables, followed by applying a filter based on the conditions \texttt{pets.pettype = 'dog'} and \texttt{student.sex = 'F'}.

Through predicate pushdown optimization, the overarching condition of the outer query is decomposed into sub-conditions that are applied directly within the inner queries. Consequently, in the modified SQL, the condition \texttt{student.sex = 'F'} is applied to the \texttt{student} table, and \texttt{pets.pettype = 'dog'} is applied to the \texttt{pets} table, prior to executing a $\theta$-join on these tables. This optimization allows for more efficient data processing by leveraging early filtering and improved index performance.

% The construction of a \textit{RelNode} by the first SQL triggers Calcite's internal predicate pushdown optimization, which consequently transforms it into the form of the second SQL. So eventually both SQLs will be built into the \textit{RelNode} structure shown in the figure \ref{relnode_eq}.

\section{False Positive Cases}

\begin{figure}[H]
    \centering
    \includegraphics[width=1\linewidth, scale=1.00]{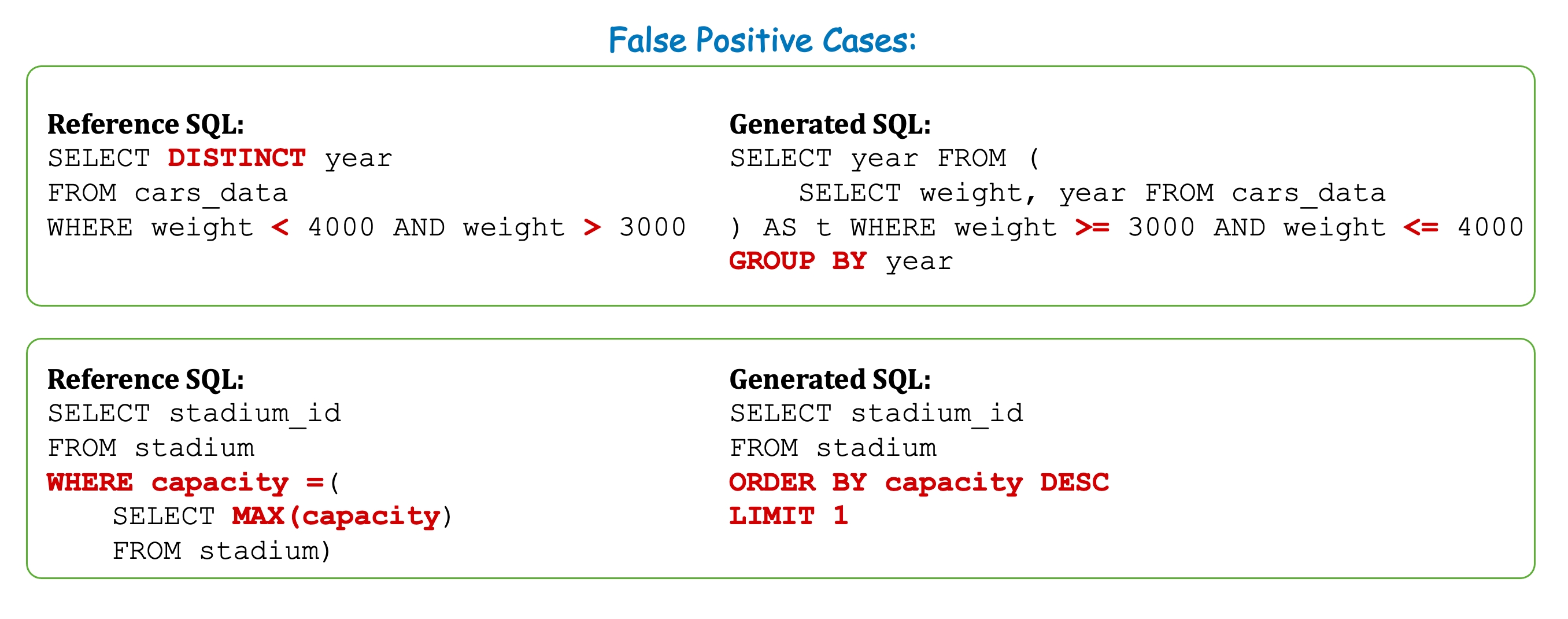}
    \caption{False positive cases. }
    \label{false_positive_cases}
\end{figure}

% In this section, we present two cases shown in Figure \ref{false_positive_cases} to discuss the issue of false positives in execution matching.

% In the first case, the sole semantic difference between the SQLs is that the filtering condition in the Reference SQL is \texttt{weight > 3000 AND weight < 4000}, whereas the Generated SQL specifies \texttt{weight >= 3000 AND weight <= 4000}. Due to the absence of any records in the database with a \texttt{weight} field exactly equal to 3000 or 4000, both SQL queries retrieve the same data, resulting in an erroneous execution match where these two non-equivalent SQL are deemed equivalent.

% In the second case, whether the two SQL statements are equivalent depends on the DDL settings. If \texttt{capacity} is set as unique, there will only be one record in the database with the maximum \texttt{capacity} value, and thus both SQL queries will yield the same single record. However, if \texttt{capacity} is not set as unique, it is possible for multiple records in the database to share the maximum \texttt{capacity} value, which would mean that the Reference SQL and Generated SQL are not equivalent. But if, by chance, there is only one record with the maximum \texttt{capacity} in the database, both SQL queries will retrieve the same data, leading to an incorrect execution match where these two non-equivalent SQL statements are deemed equivalent.

In this section, we will introduce two cases illustrated in Figure \ref{false_positive_cases} to discuss the issue of false positives in measuring the correctness of SQL queries, specifically cases where functionally incorrect SQL queries are mistakenly deemed correct.

In the first case, the sole semantic difference between the two SQL queries lies in the range specified by their filtering conditions. The reference SQL uses the conditions \texttt{weight > 3000 AND weight < 4000}, while the generated SQL specifies \texttt{weight >= 3000 AND weight <= 4000}. Since there are no records in the database with a \texttt{weight} field exactly equal to 3000 or 4000, these two SQL queries retrieve the same set of data, incorrectly treating these inequivalent SQLs as equivalent.

In the second case, whether the two SQL statements are equivalent depends on the DDL settings. If \texttt{capacity} is set to be unique, then there will only be one record in the database with the maximum \texttt{capacity} value, resulting in both SQL queries producing the same single record. However, if \texttt{capacity} is not set to unique, there could be multiple records sharing the maximum \texttt{capacity} value. This means that the reference SQL and the generated SQL are not equivalent, leading to an erroneous execution match.

\section{Spider-Pair Dataset}
\label{sec: parispider}
Text-to-SQL refers to the process of translating natural language queries into precise SQL commands \cite{yu2018spider, iyer2017learning, Deng_2021, yaghmazadeh2017sqlizer, Finegan_Dollak_2018}.
% \wh{more cite? only one is weird}. 
Numerous datasets, including Spider \cite{yu2018spider}, have been developed for this purpose, where each entry comprises a reference SQL query, a corresponding natural language question, and the relevant database. However, there is no dataset available to validate the consistency between the quality of generated SQL code and evaluation metrics. Our proposed \texttt{Spider-Pair} fills this gap. It consists of a training set and two testing sets, which we refer to as train, test, and test-aug. Each entry in the dataset includes a pair of SQL queries (reference and generated SQL), a prompt, and the functional correctness of the generated SQL. In the following sections, we will introduce our construction approach.

% However, no dataset enables a case-by-case analysis of how well code generation tasks can meet functional correctness standards. \wh{YS: We propose the dataset, on which we can validate consistency between generated SQL code quality versus an evaluation metric.}
% We propose \texttt{Spider-Pair}, a novel dataset auto-constructing pairs of SQL queries (reference and generated) and annotating them with labels indicating the equivalence of their functionality.  
% However, no dataset specifically focuses on constructing pairs of SQL queries (reference and generated) and annotating them with labels indicating the equivalence of their functionality. 
% Such a dataset would enable a case-by-case analysis of how well code generation tasks can meet functional correctness standards. 
% To our knowledge, we are the pioneers in creating a dataset of SQL query pairs labeled with functional correctness. 

\subsection{SQL Pairs Auto-generated by LLMs}
\label{sec: data SQL llm}
To generate SQL pairs, we utilize Spider \cite{yu2018spider}, as our source dataset. 
It comprises $10,181$ queries and $5,693$ unique, complex SQL queries spanning $200$ databases across $138$ distinct domains.
% LLM
We utilized $8,659$ examples from $146$ databases as our source train set, while the test set contains $1,034$ examples from $20$ different databases. 
This separation ensures fairness in evaluation by having distinct databases for train and test sets. As the capabilities of Large Language Models (LLMs) continue to advance, surpassing human performance in various tasks, we leverage them as an intermediary to generate SQLs. We carefully designed the prompts formed from the questions and the Data Definition Language (DDL) of the required databases. The DDL is crafted from the schemas of all tables within our database, where each schema outlines the table's structure in markdown format. The LLMs we used to generate SQLs includes GPT-3.5, GPT-4, and CodeLLaMA-13B.

% Before generating the corresponding SQL, it is necessary to construct prompts for the large language models (LLMs). 

\begin{table}[h]
\centering
\caption{Execution Scores of LLMs on Train and Test Sets}
\label{combined-table}
\begin{tabular}{cccc|cccc}
\hline
\multicolumn{4}{c|}{\textbf{train}} & \multicolumn{4}{c}{\textbf{test}} \\ \hline
\textbf{Model}      & \textbf{-1}   & \textbf{0}    & \textbf{1}    & \textbf{Model}      & \textbf{-1}  & \textbf{0}   & \textbf{1}   \\ \hline
\textbf{GPT3.5}     & 524  & 2722 & 5413 & \textbf{GPT3.5}     & 62  & 368 & 604 \\ 
\textbf{GPT-4}       & 212  & 2784 & 5663 & \textbf{GPT-4}       & 14  & 339 & 681 \\ 
\textbf{LLaMA}   & 1469 & 2958 & 4289 & \textbf{LLaMA}   & 165 & 361 & 508 \\ \hline
\textbf{Distinct}& 1824 & 6455 & 8769 & \textbf{Distinct}& 196 & 683 & 691 \\ \hline
\end{tabular}
\end{table}
% 这一段先这样吧
The comparative performance of three LLMs on the Spider dataset is presented in Table \ref{combined-table}. 
We categorize the output SQL statements into three ratings: -1, 0, and 1. A rating of -1 signifies the presence of syntactical mistakes in the SQL, making it non-executable by the database. 
% A rating of 0 is assigned to SQL statements that, while executable \wh{YS: meaning?}, produce results that deviate from the expected outcomes of the reference SQLs. 
% A rating of 1 indicates a perfect match in execution results with the reference. 
% Only SQL statements that the database can execute are retained for further analysis. 
% To refine our analysis, we have implemented a deduplication process for the combined results from all three models. 
% Due to a high rate of false positives in Execution Accuracy, we have conducted a thorough cleaning of the dataset. 
% To minimize the workload, we utilized the capabilities of GPT-4 for assessing the functional correctness of SQL pairs, supplemented by manual cross-verification for accuracy.
For executable SQL queries, we use 1 to indicate fidelity to the prompt instructions and 0 to indicate the opposite. Alternatively, 1 signifies equivalence in functionality to the reference SQL, while 0 indicates non-equivalence. First, we validate SQL queries based on the correctness of their execution against the database. Due to the high occurrence of false positives in execution accuracy, we then prompt GPT-4 to conduct further inspection. In cases where the execution results differ from GPT-4's response, we conduct manual evaluation to determine the functional correctness. Finally, we implement a duplicate data removal process for all SQL pairs and exclude cases where the reference and generated codes are identical.

% For SQLs which are executable, we use 1 to indicate SQLs which are faithful to instruction of the prompt and 0 which are not faithful. Or, 1 is for SQLs functionally equivalent to the reference SQL whereas 0 is for SQLs functionally inequivalent to the reference SQL. We validate the SQL correctness against the database. At the same time, due to a high false positive rate, we prompt GPT-4 to check correctness of the generated SQL. For cases where the execution result does not agree on GPT-4's response, we conduct a human evaluation to determine the functional correctness. Finally, to refine our procedure, we have implemented a deduplication process for all SQL pairs. We also eliminate cases with identical reference code and generated code.

\subsection{Data Augmentation by Equivalent Rewriting}
\label{sec: test-aug}

% In Section \ref{sec: data SQL llm}, we utilize the same LLMs to simulate the code generation process in both the train and test sets. Thus, the SQL pairs in the two sets share identical data distributions. To enhance the robustness of our evaluation, we have escalated the complexity of the test set. Specifically, we augment the test set by performing equivalent rewriting of the reference SQL. Again, we test the functional equivalence by checking their execution results.

\begin{figure}
    \centering
    \includegraphics[width=0.7\textwidth]{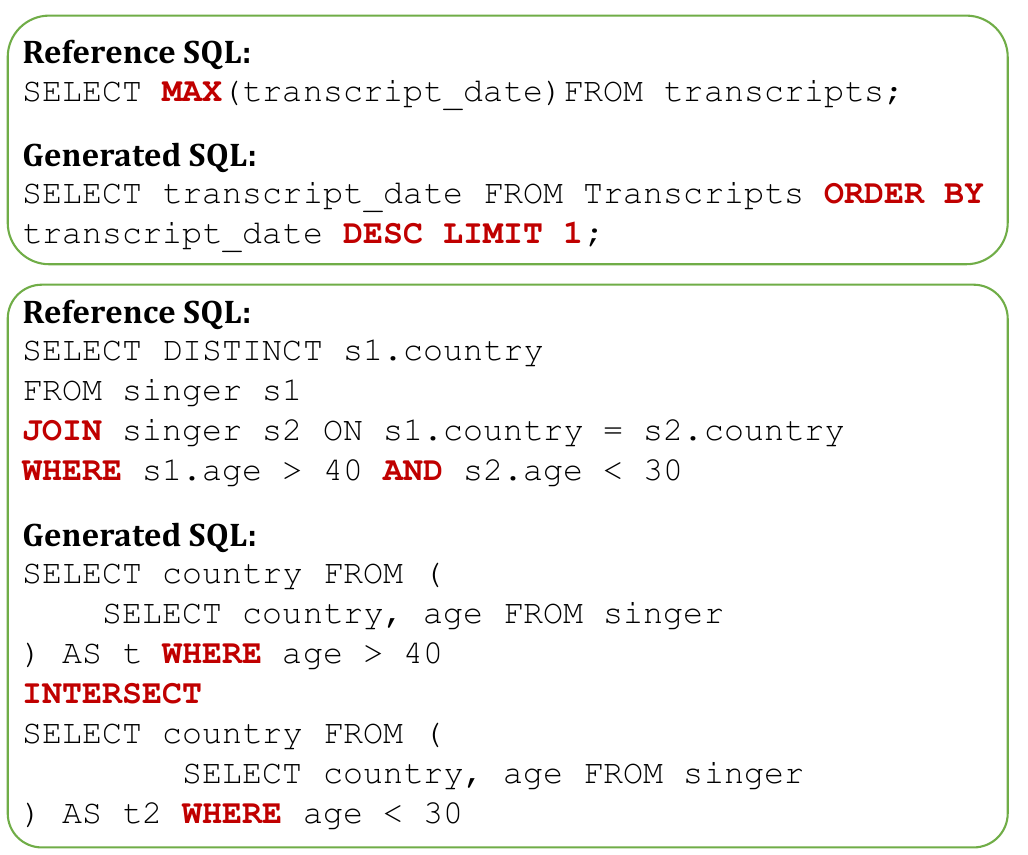}
    \caption{Hard positive SQL pairs. \textnormal{The red keywords indicate equivalent syntactic structures, where MAX can be replaced by a combination of ORDER BY and LIMIT 1. Performing a query with two conditions after joining two tables is equivalent to conducting separate conditional queries on each table and then taking the intersection.}}
    \label{hard_sql}
\end{figure}

In Section \ref{sec: data SQL llm}, we utilize the same LLMs in both the train and test sets to simulate the code generation process. Given that the same model tends to utilize similar syntactic structures to address the same type of queries, SQL pairs in both sets share identical data distributions. However, such data biases can impact the effectiveness of evaluations. To address this issue, we perform equivalence rewriting on the reference SQL, employing diverse syntactic structures to construct SQL pairs. Additionally, we verify their functional equivalence using same method in Section \ref{sec: data SQL llm}.
Figure \ref{hard_sql} illustrates two Hard Positive Cases by our equivalence rewriting, where the SQL pairs, despite achieving identical outcomes, employ significantly varied syntactic structures. 

\section{Attention Visualization For Explanation}
% In this section, we select a pair of representative and difficult SQL queries and visualize the attention maps from both the initial and final propagation to analyze the model's ability to capture feature information within the graph. 
% We then explain the process by which the GMN captures strongly associated nodes for matching by examining the changes in the attention maps between the initial and final propagation. 
% Ultimately, our analysis reveals that our model can identify equivalent substructures and functional nodes across the two graphs, thereby comprehensively understanding the structure and semantics of SQL.

% To begin with, we must clarify what constitutes a pair of SQL queries that are difficult to discern for equivalence. 
% Typically, a pair of SQL queries poses a challenge for the model when there is a significant gap between their structural representation and actual functional equivalence. 
% Following this rationale, we have selected the two SQL, which are functionally equivalent yet exhibit entirely different execution logics.

In this section, we analyze a hard positive case that two SQLs are functionally equivalent but differ in their syntactic structure. We visualize the initial and final attention graphs propagated by our \texttt{FuncEvalGMN} to examine the its capacity to identify key features within the graph. Furthermore, by observing the modifications in the attention graph from initial to final propagation, we elucidate how \texttt{FuncEvalGMN} detect and match nodes with strong correlations. This analysis demonstrates that our model is adept at recognizing equivalent substructures and functional nodes across both graphs, thereby facilitating a thorough comprehension of the SQL's syntactic structure and semantics.

\subsection{Attention Map from the First Propagation}
\begin{figure}[H]
    \centering
    \includegraphics[width=\linewidth, scale=1.00]{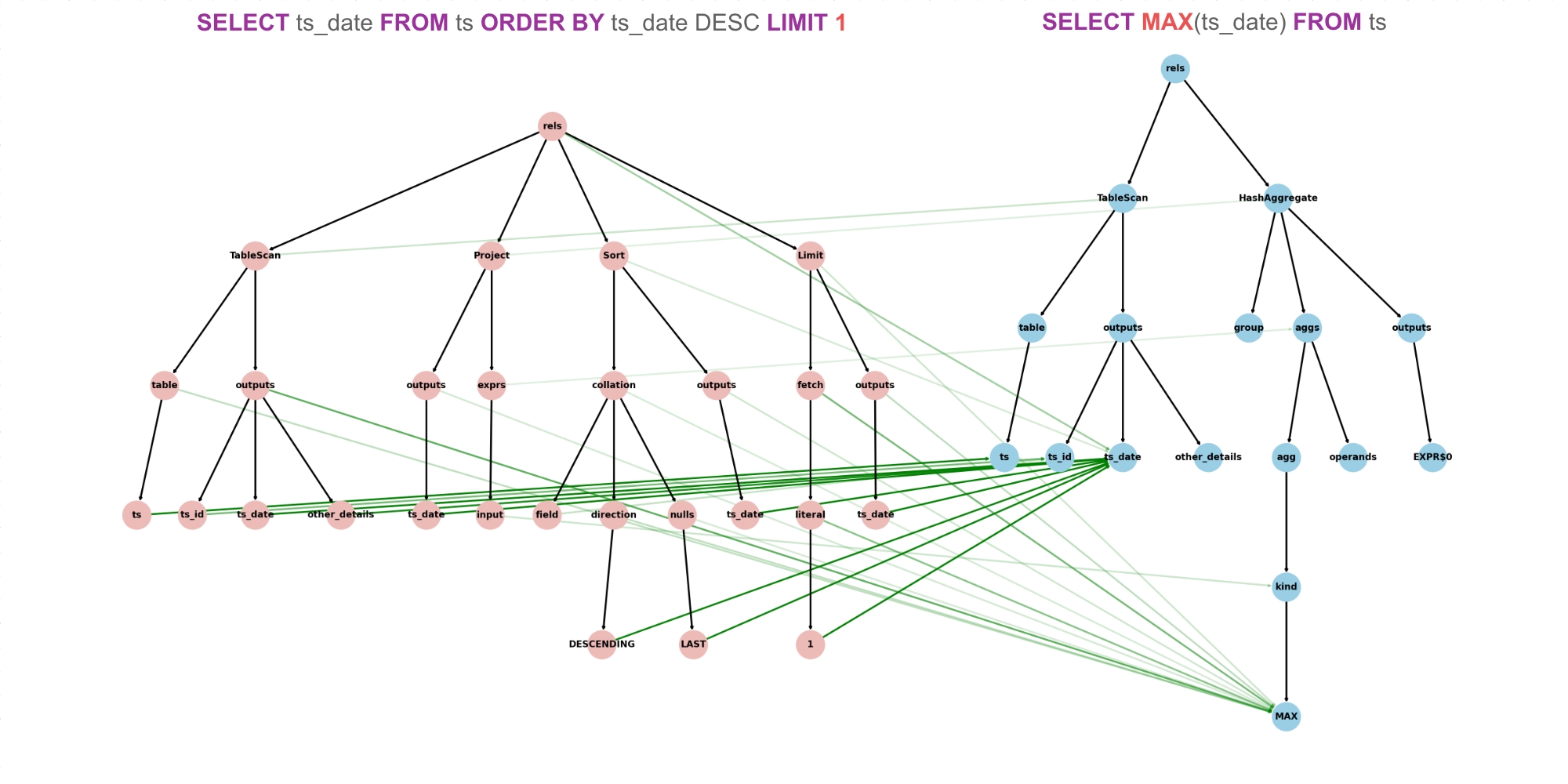}
    \caption{Attention map from the first propagation. \textnormal{In the graph, the labels on the nodes denote their content. Black edges illustrate the connections in the \textit{RelNode}, while green edges represent attention links. The transparency level of the green edges reflects the magnitude of the attention weights. Attention links are drawn from nodes in the left graph to the node that receives the highest attention out of all nodes in the right graph.}}
    \label{prop_first}
\end{figure}

% Figure \ref{prop_first} exhibits a prominent feature: the attention is predominantly focused on the \texttt{ts\_date} and \texttt{MAX} nodes. This is because \texttt{ts\_date} is the field retrieved by the SQL query, and \texttt{MAX} represents the logic to extract this field. Together, they essentially define the core functionality expressed by the SQL query. This indicates that our model has successfully captured the key feature of SQL in the right graph from the very beginning.

Figure \ref{prop_first} shows a notable feature: attention is primarily focused on the \texttt{ts\_date} and \texttt{MAX} nodes. This is because \texttt{ts\_date} is the field retrieved by the SQL query, and \texttt{MAX} is the logic used to extract it. Together, they essentially define the core functionality expressed by the SQL query. This indicates that our model has successfully captured the key semantics of SQL from the beginning.

\subsection{Attention Map From the Final Propagation}

\begin{figure} [H]
    \centering
    \includegraphics[width=\linewidth, scale=1.00]{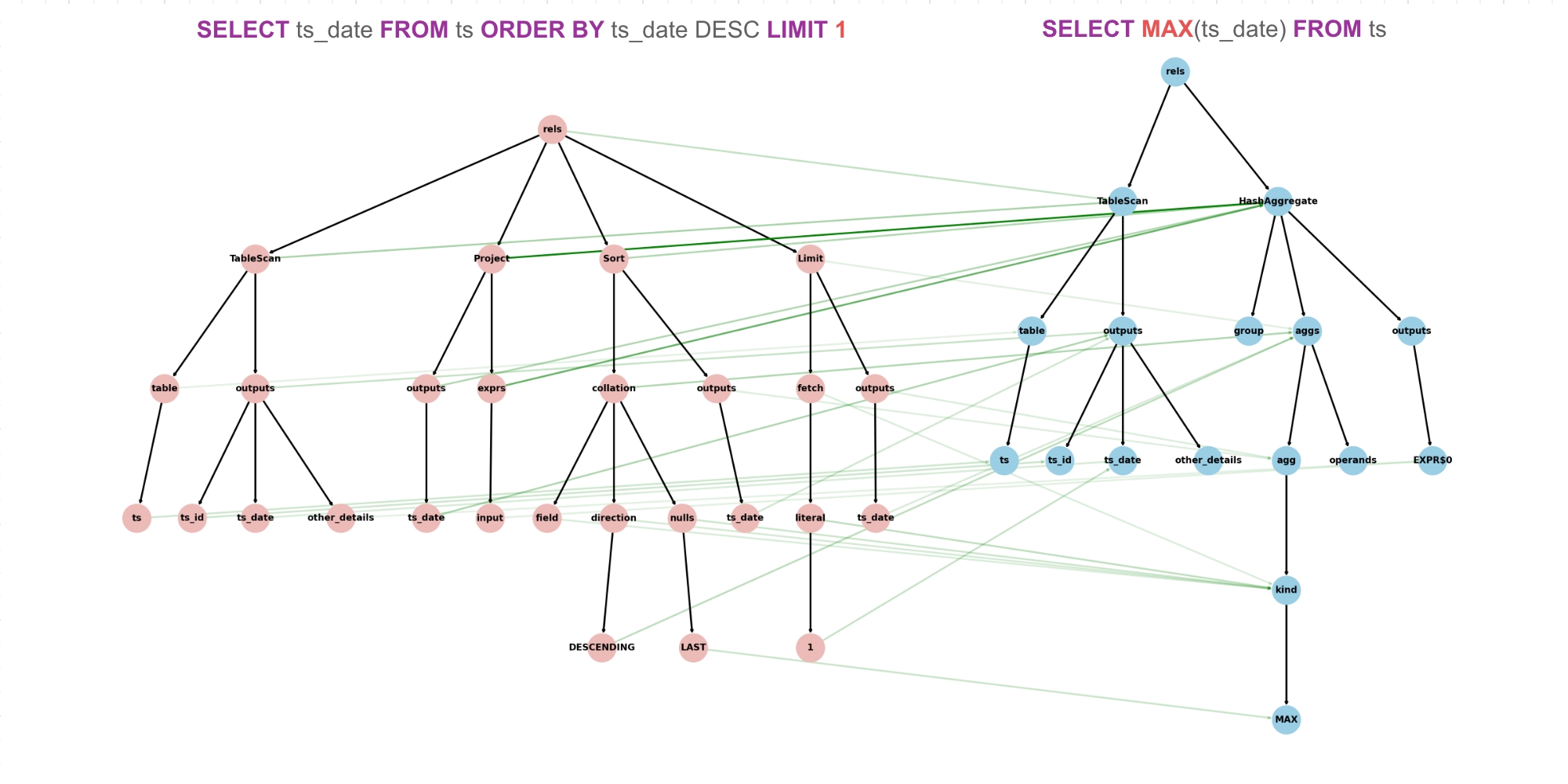}
    \caption{Attention map from the final propagation. }
    \label{prop_final}
\end{figure}

Figure \ref{prop_final} displays the attention map from the model's final propagation, where we can identify several key features:

% From the perspective of relational operations, the presence of the \texttt{TableScan} subtree in both graphs indicates that they share the same \texttt{TableScan} operation.

\begin{enumerate}
    \item \textbf{Equivalent Substructures Captured}\\
    It is observed that all corresponding nodes of the \texttt{TableScan} subtree, except for the \texttt{other\_details} node, are successfully matched by attention edges in both graphs. This phenomenon can be explained from two perspectives:
    \begin{enumerate}
        \item[a.] In terms of similarity, the \texttt{TableScan} subtree, being a common element in both SQL queries, exhibits the highest degree of similarity.
        \item[b.] Furthermore, nearly all nodes within the \texttt{Project}, \texttt{Sort}, and \texttt{Limit} subtree in the left graph are matched with attention edges to the \texttt{HashAggregate} subtree in the right graph. This is because the combination of \texttt{Project}, \texttt{Sort}, and \texttt{Limit} operations in the left graph is functionally equivalent to the \texttt{HashAggregate} operation in the right graph. From this analysis, it is evident that our model possesses a strong capability to extract equivalent functional structures from entirely different structures.
    \end{enumerate}

    \item \textbf{Equivalent Functional Nodes Captured}\\
    Observation reveals that the \texttt{LAST} node in the \texttt{Sort} subtree of the left graph draws attention to the \texttt{MAX} node in the right graph. This is because the operation of sorting in descending order and selecting the last data entry is equivalent to directly taking the \texttt{MAX} in an aggregate function. Additionally, \texttt{ts\_date}, as the field resulting from the SQL execution, is precisely captured: the \texttt{1} node within the \texttt{limit} subtree in the left graph draws attention to the \texttt{ts\_date} node in the right graph. This occurs because using \texttt{limit} to retrieve the last record extracts the \texttt{ts\_date} field, and this node, being the only \texttt{ts\_date} node in the right graph, is accurately identified.
\end{enumerate}

\subsection{Attention Map Comparison}

In the following, we compare the attention maps from the initial and final propagations to analyze the trends in the attention map changes throughout the model's propagation process, ultimately discerning the capabilities and characteristics of the model's attention component in feature extraction.

Initially, the node embeddings have only been processed by the encoder layer and have not yet integrated neighborhood information and structural features. 
At this stage, the model quickly captures the SQL's core features, \texttt{ts\_date} and \texttt{MAX}, but overlooks other SQL details. 
However, after the final propagation, by observing the direction and opacity of the attention edges, we can discern that the attention map exhibits the following four characteristics:

\begin{enumerate}
    \item The distribution of attention is more uniform.
    \item The attention weights are more balanced.
    \item Equivalent substructures within the two graphs are captured.
    \item Functionally equivalent nodes across different structures are identified.
\end{enumerate}

These observations indicate that as propagation progresses, the model begins to consider a broader range of features within the SQL graph, moving beyond the initial focus on key elements to a more comprehensive understanding of the SQL's structure and semantics. Additionally, the model is capable of extracting functionally consistent information from both equivalent functional nodes and equivalent substructures. 
This capability is, to some extent, due to the use of the relational operator tree (ROT), as each subtree (substructure) within the ROT represents an atomic functional operation in the execution plan.

\section{Test Set AUC Trends During Training}
\label{auc: appendix}

\begin{figure}[H]
  \begin{minipage}[t]{0.5\textwidth}
    \centering
    \includegraphics[width=\textwidth]{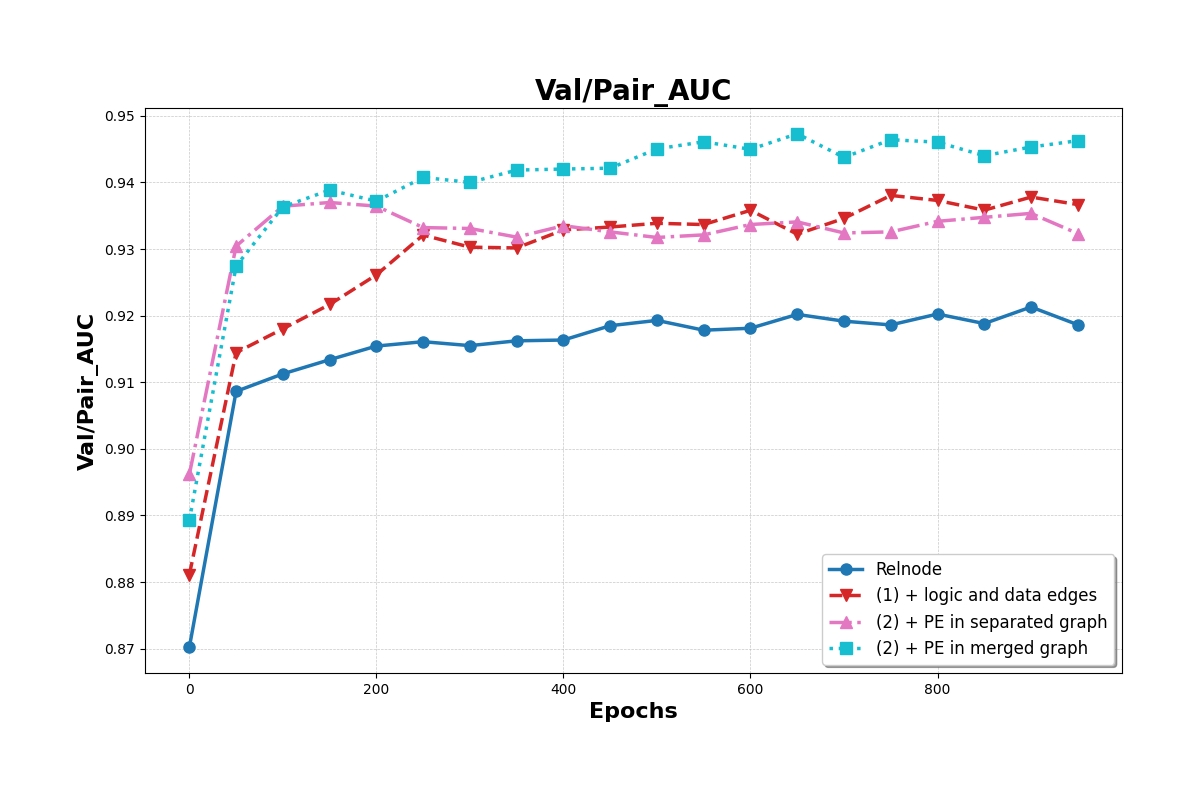}
    \caption{AUC in test dataset}
    \label{test_auc}
  \end{minipage}\hfill
  \begin{minipage}[t]{0.5\textwidth}
    \centering
    \includegraphics[width=\textwidth]{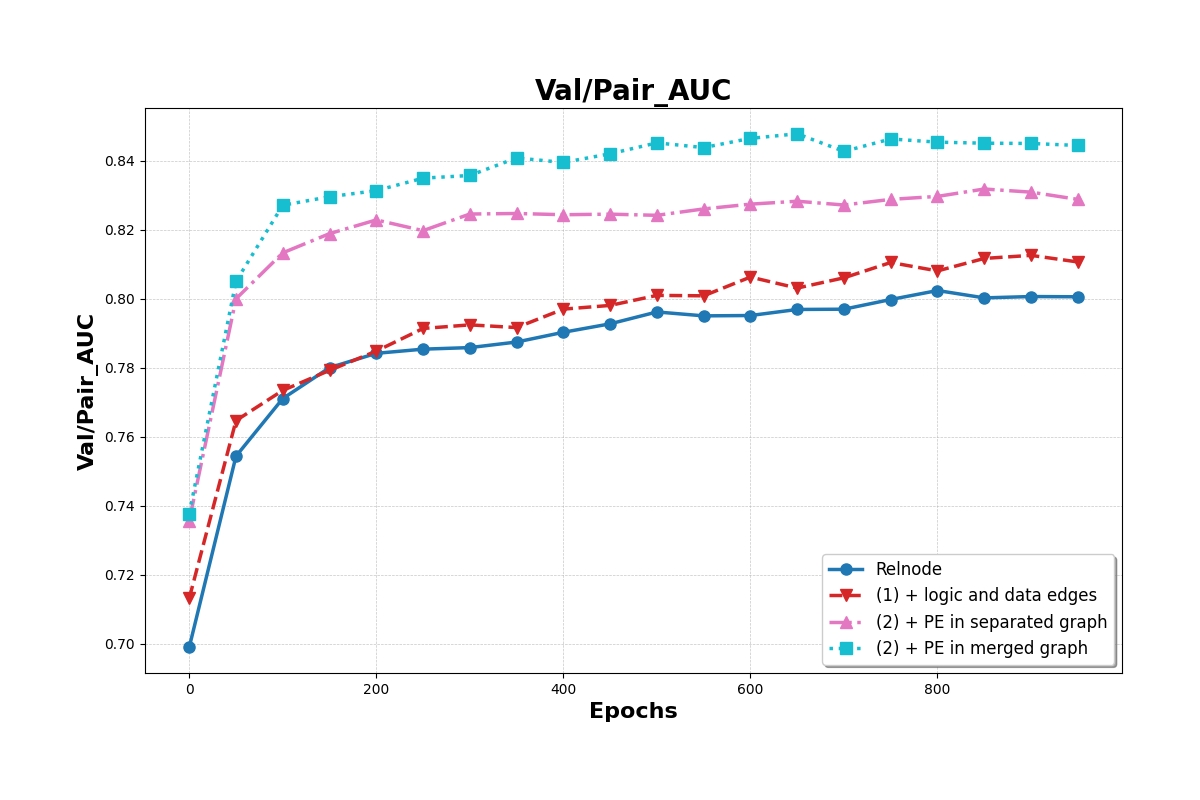}
    \caption{AUC in test\_aug dataset}
    \label{test_aug_auc}
  \end{minipage}
\end{figure}

\section{Correlation Evaluation}
\label{a: correlation evaluation}
The performance of different models can be evaluated using the following metrics:

\textbf{Area Under the Curve (AUC)}\citep{huang2005using} refers to the area under the Receiver Operating Characteristic (ROC) curve. The ROC curve is a graphical representation that illustrates the diagnostic ability of a binary classifier system as its discrimination threshold is varied.

$$AUC =\int_{}^{}\mathrm{TPR}(\xi)\mathrm{FPR}^{\prime}(\xi)\mathrm{~}d\xi$$ where \( TPR(\xi) \) is the true positive rate and \( FPR(\xi) \) is the false positive rate at threshold \( \xi \), and \( FPR'(\xi) \) takes the differentiation with respect to the threshold. 

% is a non-parametric measure of rank correlation (statistical dependence between the rankings of two data):

% where $\operatorname {R} (M^1)$ and $\operatorname {R} (M^2)$ represent the rankings of $M^1$ and $M^2$, ${\displaystyle \operatorname {cov}(\cdot, \cdot)}$ means the covariance function \wh{YS: what is a covariance function?}, and ${\displaystyle \sigma_{M}}$ means the standard deviation of $M$.

% This measurement evaluates the degree to which the ranking order produced by an evaluation metric aligns with that of a reference ranking, serving as a tool to understand the similarity between two sets of ranked data.
\textbf{Spearman R ($r_s$)}\citep{pranklin1974introduction} is a nonparametric measure of rank correlation, which assesses the statistical dependence between the rankings of two variables or data sets:

\[
r_{s}=\frac{\operatorname{cov}(\mathrm{R}(Y^1), \mathrm{R}(Y^2))}{\sigma_{\mathrm{R}(Y^1)} \sigma_{\mathrm{R}(Y^2)}},
\]where $\operatorname{cov}(\mathrm{R}(Y^1), \mathrm{R}(Y^2))$ expresses the covariance between the rankings of $Y^1$ and $Y^2$, represented by  $\operatorname {R} (Y^1)$  and  $\operatorname {R} (Y^2)$ and ${\displaystyle \sigma}$ refers to the standard deviation.

% \textbf{Pearson R$\mathbf{(r_p)}$}\citep{bravais1844analyse} is a measure of linear correlation between two data:
% \[
% r_{p}=\frac{\operatorname{cov}(M^1, M^2)}{\sigma_{M^1} \sigma_{M^2}}.
% \]

% coefficient is a statistical measure that quantifies the strength and direction of the ordinal association between two variables, such as a metric (e.g., CodeBERTScore) and a reference measurement. It assesses how well the order between two rankings matches.

% \textbf{Kendall-Tau$\mathbf{(\tau)}$}\citep{kendall1938new} is a statistical metric used to measure the ordinal association between two measured data:
% \[
% \tau=\frac{Concordant - Discordant}{Concordant + Discordant},
% \]
% where $Concordant$ indicates the number of occurrences that two evaluation data $M^1$ and $M^2$ exist either both $M^1_{i}>M^1_{j}$ and $M^2_{i}>M^2_{j}$ or both $M^1_{i}<M^1_{j}$ and $M^2_{i}<M^2_{j}$, and $Discordant$ indicates the number of occurrences opposite to $Concordant$.

\textbf{Kendall-Tau ($\tau$)}\citep{kendall1938new}  assesses the relationship between two rankings by measuring the ordinal or rank correlation between a given variable and a reference measurement. The formula is:
\[
\tau = \frac{Concordant - Discordant}{Concordant + Discordant},
\] where \(Concordant\) is the number of pairs for which the two measurements agree on their relative rank. Conversely, \(Discordant\) counts the pairs in which the two measurements demonstrate conflicting ranking orders.

% \textbf{Kendall-Tau$\mathbf{(\tau)}$}\citep{kendall1938new}  evaluates the ordinal or rank correlation between a variable and a benchmark measurement. It is defined by the equation:

% \[
% \tau=\frac{Concordant - Discordant}{Concordant + Discordant},
% \]

% where \(Concordant\) denotes the count of concordant pairs, where two sets of measurements concur on the order of their rankings. Conversely, \(Discordant\) signifies the count of discordant pairs, where the two sets of measurements disagree on the order.

 % Specifically, if \(g(x_1, z_1) > g(x_2, z_2)\), then the benchmark also indicates \(g'(x_1, z_1) > g'(x_2, z_2)\). 

% \subsection{Ablation experiment}
% 1. Different AST construction methods: \textit{RelNode}, simplified \textit{RelNode}, ast, different data flow \\
% 2. Different Node Embedding: 1d-Conv, resnet, node mapping(tianyu's idea, todo), hash\\
% 3. Different GMN matching methods: different pooling, updating layers, cross attention\\

\section{Evaluation of Code LMs}
Our training and testing datasets are sourced from GPT3.5, GPT-4, and CodeLlama. To validate the effectiveness of our FuncEvalGMN against other models, we conducted inference using the DeepSeek model on the Spider dataset. The obtained AUC, $r_s$, and $r_p$ are 91.64\%, 51.55\%, and 59.94\%, respectively, demonstrating that our approach also exhibits strong evaluation capabilities on other large models.

\begin{figure}[H]
    \centering
    \includegraphics[width=1\linewidth, scale=1.00]{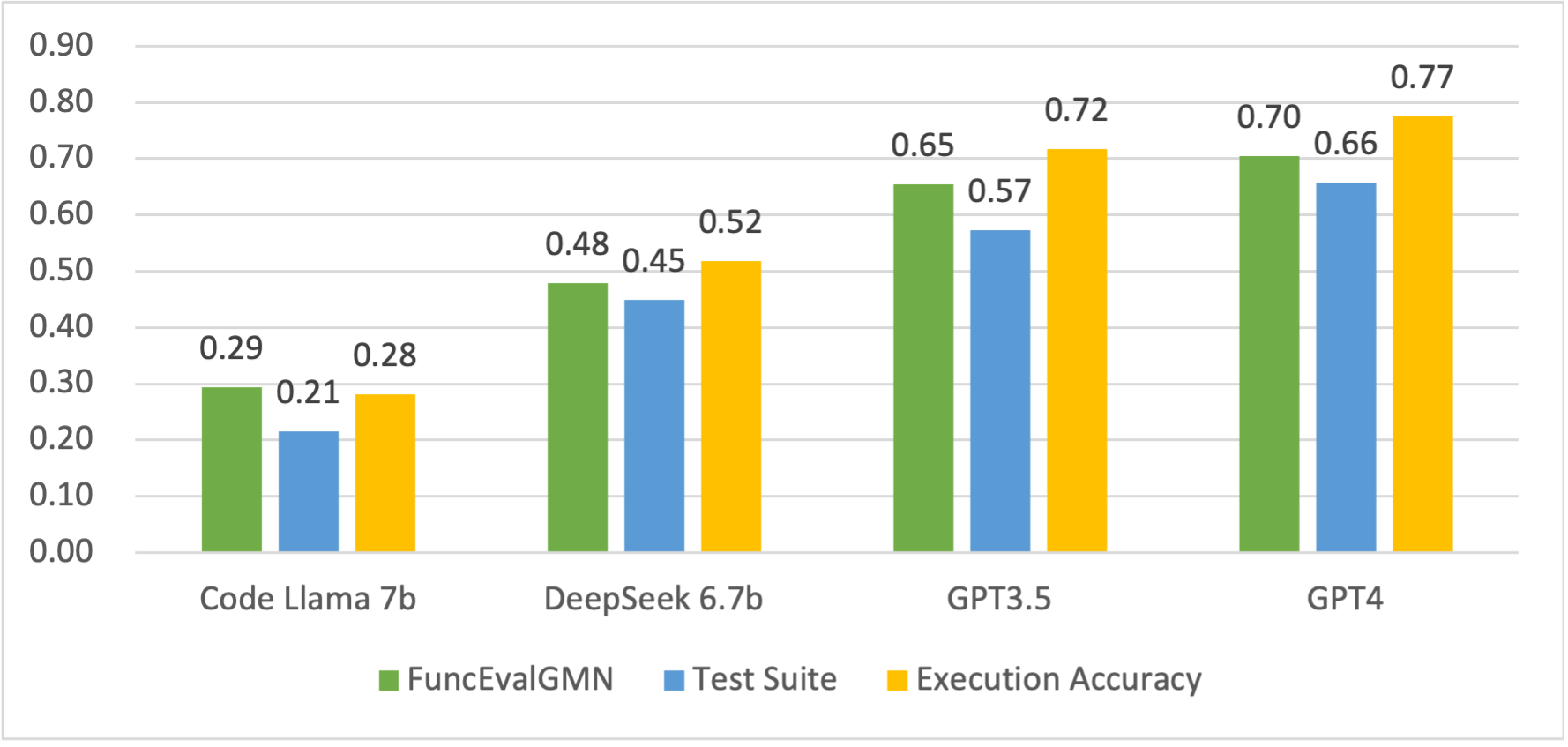}
    \caption{Evaluation of Code LMs}
    \label{codelm}
\end{figure}

Finally, we evaluated four Code Large Models (LMs) on the Spider dataset using three evaluation metrics: \texttt{FuncEvalGMN}, Test Suite \citep{zhong2020semantic}, and Execution Accuracy \cite{yu2018spider}. The original output range of \texttt{FuncEvalGMN}, denoted as \( y \), ranged from negative infinity to zero. We normalized the output results to the range of 0 to 1 using the formula \( y = \max\left[\frac{y + 3}{3}, 0\right] \). As shown in Figure \ref{codelm}, our \texttt{FuncEvalGMN} can also serve as a good metric for evaluating SQL generation. Compared to Execution Accuracy and Test Suite, we do not incur the cost of maintaining and executing databases.

\section{Keywords Distribution in BIRD and SPIDER Dataset}
\label{sec:keyword_distribution}

In our study, we analyzed two datasets, BIRD and SPIDER, to understand the distribution of SQL keywords. This analysis provides insight into the complexity and nature of queries present in each dataset.

The BIRD dataset exhibits a high frequency of the \texttt{WHERE} keyword, present in 90.29\% of the queries. This indicates a strong emphasis on filtering conditions in the queries. \texttt{JOIN} operations are prevalent, appearing in 74.32\% of the queries, suggesting a significant number of queries involve combining data from multiple tables.

The SPIDER dataset presents a different distribution of SQL keywords. The \texttt{WHERE} keyword is used in 47.68\% of the queries, significantly less frequent than in the BIRD dataset, indicating fewer conditions are applied to filter data. The \texttt{JOIN} keyword appears in 39.46\% of the queries, showing less reliance on combining tables compared to BIRD.

Interestingly, \texttt{Aggregation} (53.29\%) and \texttt{Counting} (39.85\%) are more common in SPIDER, suggesting a higher focus on data summarization. The \texttt{Order By} keyword is used in 22.34\% of the queries, and \texttt{LIMIT} appears in 17.70\%, both slightly higher than in BIRD.

The analysis of these datasets reveals distinct patterns in SQL keyword usage, reflecting the different types of queries they encompass. BIRD has a higher emphasis on filtering and joining data, while SPIDER shows a greater focus on data aggregation and union operations. These differences highlight the varied query complexities and use cases catered to by each dataset.

\begin{table}[h]
    \centering
    \begin{tabular}{lcc}
        \toprule
        Keyword       & BIRD (\%)       & Spider (\%)    \\
        \midrule
        Where         & 90.29           & 47.68          \\
        Join          & 74.32           & 39.46          \\
        Aggregation   & 42.96           & 53.29          \\
        Counting      & 31.29           & 39.85          \\
        Order By      & 18.64           & 22.34          \\
        Limit         & 18.25           & 17.70          \\
        Distinct      & 15.45           & 8.41           \\
        StrFunc       & 13.49           & N/A            \\
        Cast          & 10.04           & N/A            \\
        Average       & 3.59            & 6.48           \\
        Subquery      & 3.59            & 4.84           \\
        MinMax        & 2.48            & 5.71           \\
        Sum           & 2.67            & 2.61           \\
        Union         & 0.13            & 7.74           \\
        \bottomrule
    \end{tabular}
    \caption{Performance of Keywords in BIRD and Spider}
    \label{tab:keyword_performance}
\end{table}

\end{document}